\documentclass[prl,twocolumn,preprintnumbers,nofootinbib,tightenlines,superscriptaddress]{revtex4}
\usepackage{float}
\usepackage{graphicx}
\usepackage{amssymb}
\usepackage{amsmath,mathrsfs,verbatim}
\usepackage{times}
\usepackage{latexsym}
\usepackage{soul}
\def\beq{\begin{equation}}
\def\eeq{\end{equation}}
\def\bey{\begin{eqnarray}}
\def\eey{\end{eqnarray}}

\def\kpc{\, {\rm kpc} }
\def\msun{M_\odot}

\def\lsim{\mathrel{\raise.3ex\hbox{$<$\kern-.75em\lower1ex\hbox{$\sim$}}}}
\def\gsim{\mathrel{\raise.3ex\hbox{$>$\kern-.75em\lower1ex\hbox{$\sim$}}}}

\newcommand{\be}{\begin{equation}}
\newcommand{\ee}{\end{equation}}

\newcommand{\vmax}{V_{\rm max}}

\newcommand{\vc}{V_{\rm cir}}
\newcounter{sec}

\newcommand{\sdm}{\Sigma_{\rm DM,0}}

\begin{document}

\title{How the Self-Interacting Dark Matter Model Explains the Diverse Galactic Rotation Curves}
\author{Ayuki Kamada}
\affiliation{Department of Physics and Astronomy, University of California, Riverside, California 92521, USA}
\affiliation{Institute for Basic Science, Center for Theoretical Physics of the Universe, Daejeon 34051, South Korea}
\author{Manoj Kaplinghat}
\affiliation{Department of Physics and Astronomy, University of California, Irvine, California 92697, USA}
\author{Andrew B. Pace}
\affiliation{Department of Physics and Astronomy, University of California, Irvine, California 92697, USA}
\affiliation{George P. and Cynthia Woods Mitchell Institute for Fundamental Physics and Astronomy, and Department of Physics and Astronomy, Texas A\&M University, College Station, TX 77843, USA}
\author{Hai-Bo Yu}
\affiliation{Department of Physics and Astronomy, University of California, Riverside, California 92521, USA}
\date{\today}
\begin{abstract}
\vspace*{.0in}

The rotation curves of spiral galaxies exhibit a diversity that has been difficult to understand in the cold dark matter (CDM) paradigm. 
We show that the self-interacting dark matter (SIDM) model provides excellent fits to the rotation curves of a sample of galaxies with asymptotic velocities in the 25 to 300 km/s range that exemplify the full range of diversity. 
We only assume the halo concentration-mass relation predicted by the CDM model and a fixed value of the self-interaction cross section.
In dark matter dominated galaxies, thermalization due to self-interactions creates large cores and reduces dark matter densities. 
In contrast, thermalization leads to denser and smaller cores in more luminous galaxies, and naturally explains the flat rotation curves of the highly luminous galaxies. 
Our results demonstrate that the impact of the baryons on the SIDM halo profile and the scatter from the assembly history of halos as encoded in the concentration-mass relation can explain the diverse rotation curves of spiral galaxies. 

\end{abstract}
\maketitle

\stepcounter{sec}
{\bf \Roman{sec}. Introduction.\;} The $\Lambda$CDM model, with a cosmological constant ($\Lambda$) and cold dark matter (CDM), explains the observed large-scale structure of the Universe~\cite{Ade:2013zuv} and many aspects of galaxy formation~\cite{Springel:2006vs,TrujilloGomez:2010yh}, but the diverse observed rotation curves do not have a satisfactory explanation. Observations of a number of dwarf and low surface brightness galaxies indicate that the inner halo is often badly fit by the cusped halos predicted by $\Lambda$CDM simulations~\cite{Flores:1994gz,Moore:1994yx,  Burkert:1995yz,Persic:1995ru, deBlok:2002tg, Gentile:2004tb, KuziodeNaray:2007qi, deblok2008, Oh:2010ea,Oh:2015xoa}. The core densities exhibit almost an order of magnitude spread for similar total halo masses~\cite{deNaray:2009xj}, and galaxies with densities at the upper end of the range are consistent with $\Lambda$CDM~\cite{Oman:2015xda}. There is no clear explanation for the diversity in the inner rotation velocity profiles of different galaxies within similar mass halos~\cite{Oman:2015xda}. 

In this {\it Letter}, we demonstrate how this diversity problem~\cite{Oman:2015xda}
can be solved in the self-interacting dark matter (SIDM) framework~\cite{Spergel:1999mh}, where dark matter particles exchange energy by colliding with one another in halos. Dark matter self-interactions only change the inner halo properties in accord with observations, leaving all the successes of CDM intact on large scales, e.g.,~\cite{Vogelsberger:2012ku,Rocha:2012jg,Peter:2012jh,Vogelsberger:2015gpr}. While the original SIDM model~\cite{Spergel:1999mh} posited a cross section that is independent of velocity, constraints from galaxy clusters demand an interaction cross section that diminishes with increasing velocity~\cite{Firmani:2000ce,Yoshida:2000uw,Peter:2012jh,Kaplinghat:2015aga,Elbert:2016dbb}. SIDM models based on a Yukawa potential~\cite{Feng:2009hw,Buckley:2009in,Loeb:2010gj,Tulin:2012wi,Tulin:2013teo,Cline:2013zca,Boddy:2014yra,Schutz:2014nka} or atomic scattering~\cite{Cline:2013pca} can naturally accommodate the required velocity dependence. Concrete particle physics realizations of such models can provide a universal solution to small-scale issues from dwarf galaxies to galaxy clusters~\cite{Kaplinghat:2015aga,Boddy:2016bbu}.

The diversity in the observed rotation curves is solved by a combination of interconnected features in the $\Lambda$SIDM model. In the outer parts of galaxies, the $\Lambda$SIDM model is the same as the $\Lambda$CDM model, inheriting all its successes. In the inner regions, the SIDM density profile and its relation to the baryons is changed by the process of thermalization due to the self-interactions. The physical effects of thermalization in the inner region are varied but fully determined by the distribution of the baryons, up to the scatter from the assembly history. In many galaxies, thermalization forces particles out of the center and leads to a lower circular velocity than the dark matter-only $\Lambda$CDM predictions. 
In other galaxies where stars dominate the gravitational potential, the SIDM halo profile can be as steep as the $\Lambda$CDM predictions. In these galaxies, the total rotation curve is forced to be flat even at radii much smaller than the scale radius of the dark matter halo, providing a natural explanation to the disk-halo conspiracy~\cite{1986RSPTA.320..447V}. 

All of the features discussed above are captured in a simple model that we discuss next. We discuss the scatter due to halo concentration and the baryon distribution in Sec. III. In Sec. IV, we provide fits to representative galaxies. We discuss the uniformity in the SIDM fits and the self-scattering cross section in Secs. V and VI, respectively. We conclude in Sec. VII.

\begin{figure}[!t]
\centering
\begin{tabular}{c}
\includegraphics[trim=2.5cm 2.5cm 2.5cm 2.5cm, clip=true,scale=0.53]{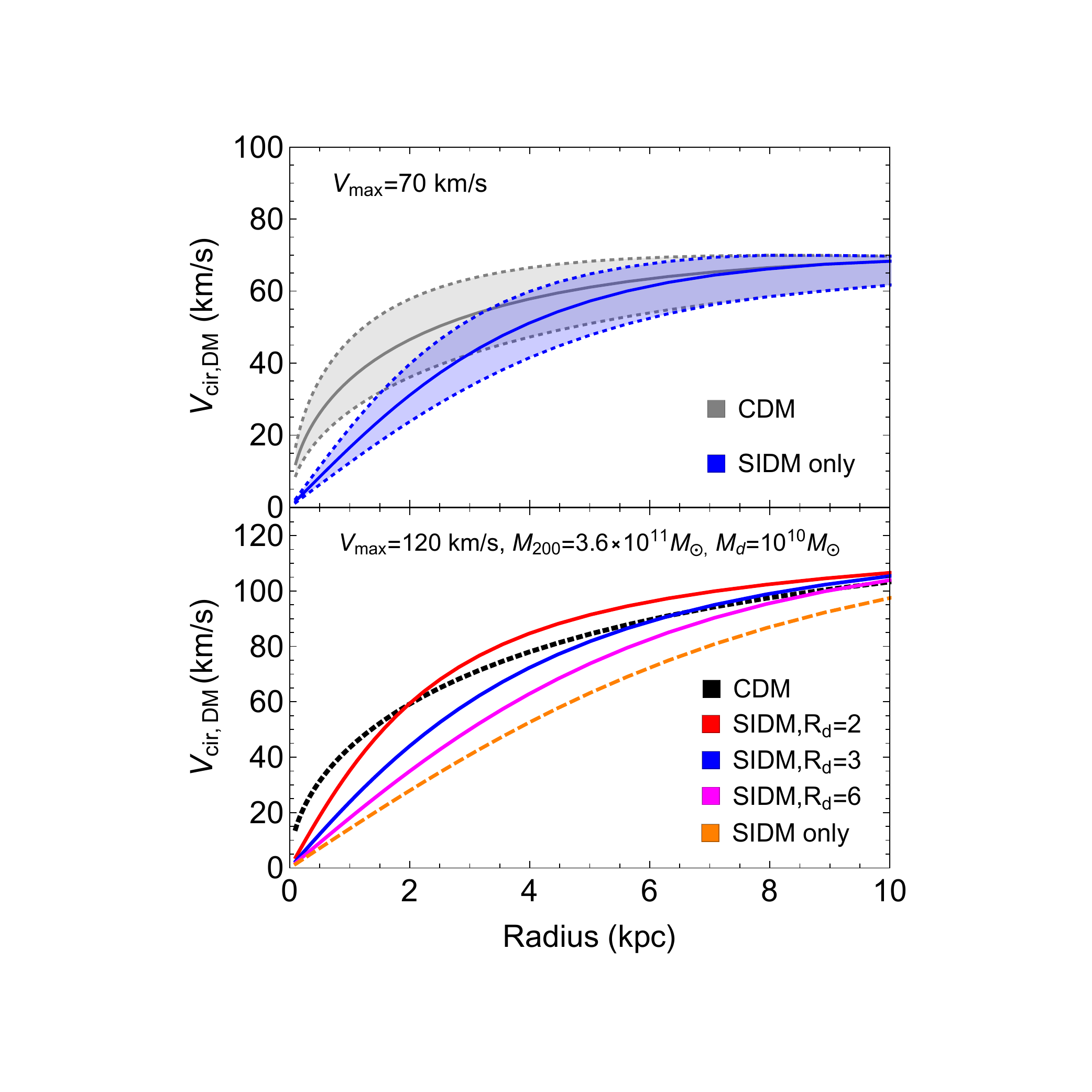}
\end{tabular}
\caption{\label{fig:sidmgeneric}{\em Top}: Circular velocity of the SIDM halo and the corresponding CDM halo for $V_{\rm max}=70~{\rm km/s}$ with the $2\sigma$ spread in halo concentration. {\em Bottom}: Circular velocity of the SIDM halo with $V_{\rm max}=120~{\rm km/s}$ and median concentration including the impact of a stellar disk of mass $M_d=10^{10}~{M_\odot}$ for three disk scale lengths $R_d=2~{\rm kpc}$, $3~{\rm kpc}$, and $6~{\rm kpc}$. The corresponding SIDM (dashed) and CDM (dotted) circular velocities without disks are also shown.}
\end{figure}

\stepcounter{sec}
{\bf \Roman{sec}. Modeling the SIDM halo with a stellar disk.\;}We have developed an analytical method to model the SIDM halo properties~\cite{Kaplinghat:2013xca,Kaplinghat:2015aga}, which is based on the isothermal solutions to the Jeans equations. The spherically symmetric version of the model described below has been tested against cosmological N-body simulations~\cite{Kaplinghat:2015aga} and isolated SIDM simulations of a range of galaxy types including baryons~\cite{Elbert:2016dbb}.  

We divide the halo into two regions, separated by a characteristic radius $r_1$ where the average scattering rate per particle times the halo age is equal to unity. The value of $r_1$ is determined by the condition $\langle \sigma v\rangle\rho(r_1) t_{\rm age}/m\approx 1$, where $\sigma$ is the scattering cross section, $m$ is the dark matter particle mass,  $v$ is the relative velocity between dark matter particles, and  $\langle ... \rangle$ denotes averaging over the velocity distribution. In this {\em Letter}, we assume $\sigma/m=3~{\rm cm^2/g}$ and $t_{\rm age}=10$ Gyr for all the galaxies motivated by previous results~\cite{Kaplinghat:2015aga}. 

For radii $r < r_1$, SIDM particles experience multiple collisions over the age of galaxies and reach kinetic equilibrium. The density profile from integrating out the velocities in the Maxwellian phase space distribution is $\rho_{\rm iso}(\vec{r})=\rho_0\exp[-\Phi_{\rm tot}(\vec{r})/\sigma^2_{\rm v0}]$, 
where $\rho_0$ is the central dark matter density, $\sigma_{\rm v0}$ is the one dimensional dark matter velocity dispersion, and $\Phi_{\rm tot}$ is the total gravitational potential due to dark and baryonic matter normalized such that $\Phi_{\rm tot}(0)=0$. We note that features in the stellar or gas potential get imprinted in $\rho_{\rm iso}$ through $\Phi_{\rm tot}$; while we do not model such baryonic features here, we expect they will be more {\it strongly} reflected in the rotation curve~\citep{Sancisi2004} in SIDM than CDM. In addition, the impact of the baryons on the SIDM halo profile is determined by its gravitational potential relative to $\sigma^2_{\rm v0}$.

Assuming that the baryons are distributed in a thin disk with central surface density $\Sigma_0$ and scale radius $R_d$, we can write the Poisson equation for $\Phi_{\rm tot}$ as: 
\beq
\boldsymbol{\nabla}^2\Phi_{\rm tot}(R,z)= 4\pi G [\rho_{\rm iso}(R,z)+\Sigma_0 e^{-R/R_d}\delta(z)]\,.
\label{eq:poisson}
\eeq
We solve Eq.~(\ref{eq:poisson}) by expanding it in the Legendre polynomials~\cite{Amorisco:2010sb}. 
The solution can be parametrized by two dimensionless parameters defined as $a\equiv 8\pi G\rho_0R^2_d/(2\sigma^2_{\rm v0})$ and $b\equiv8\pi G\Sigma_0 R_d/(2\sigma^2_{\rm v0})$~\cite{Amorisco:2010sb}. We have calculated $300$ templates in total with different combinations of $a$ and $b$ values and interpolated between them as required. 

For $r > r_1$, where scattering has occurred less than once per particle on average, we model the dark matter density as the Navarro-Frenk-White (NFW) profile $\rho_{\rm NFW}(r)= \rho_s (r/r_s)^{-1}(1+r/r_s)^{-2}$ seen in $\Lambda$CDM simulations~\cite{Navarro:1996gj}. We create the SIDM profile by joining the spherically-averaged isothermal ($\rho_{\rm iso}$) and the spherical NFW ($\rho_{\rm NFW}$) profiles at $r=r_1$ such that the mass and density are continuous at $r_1$. For $\sigma/m={\cal O}(1)~{\rm cm^2/g}$, the matching implies that $r_1$ is close to $r_s$~\cite{Rocha:2012jg}. The SIDM halo parameters ($\rho_0,\sigma_{\rm v0}$) directly map onto $(r_s, \rho_s)$ or $(M_{200}, c_{200})$ of the NFW profile for a fixed $\sigma/m$. We assume that the large-scale structure is the same as that in the Planck $\Lambda$CDM model~\cite{Ade:2013zuv} and impose its halo concentration-mass relation on our solutions, $c_{200}=10^{0.905\pm0.11}\left(M_{200}/10^{12}h^{-1}M_\odot\right)^{-0.101}$~\cite{Dutton:2014xda}, while allowing for the expected scatter $0.11$ dex scatter ($1\sigma$).  

\begin{figure*}[thb]
\centering
\begin{tabular}{c}
\includegraphics[scale=1.1]{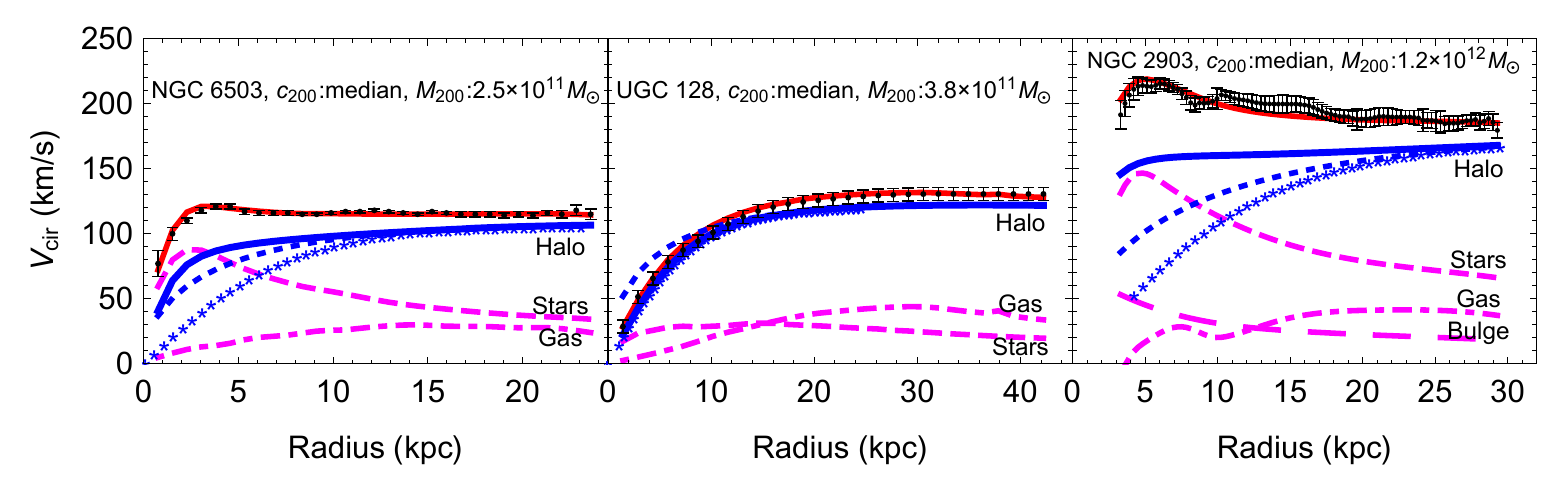}
\end{tabular}
\caption{\label{fig:examplesv130} Left two panels show the SIDM fits to the rotation curves of NGC 6503 and UGC 128. They asymptote to $V_f\approx 130~{\rm km/s}$ in the outer parts, but their inner rotation curves are very different. The right panel show the SIDM fit to a highly luminous galaxy, NGC 2903. The total fit is displayed in red and it includes contributions from the SIDM halo (blue solid), stars (magenta dashed), gas (magenta dot-dashed), and bulge (magenta long-dashed). The predictions of the corresponding CDM halo (dotted) and the SIDM halo neglecting the influence of the baryons (asterisk) are also shown.}
\end{figure*}

\stepcounter{sec}
{\bf \Roman{sec}. Halo concentration and the role of baryons.\;}In the top panel of Fig.~\ref{fig:sidmgeneric}, we show the circular velocity due to the dark matter halo, $V_{\rm cir, DM}(r)$, as a function of the radius for the SIDM and the corresponding CDM halos for $V_{\rm max}=70~{\rm km/s}$. It is clear that dark matter self-interactions can lower circular velocity systematically in the inner regions. To assess the scatter quantitatively, we use $V_{\rm cir}(2~{\rm kpc})$~\cite{Oman:2015xda}. At large radii $r\gtrsim r_1$, both halos have the same $V_{\rm cir, DM}(r)$, but the SIDM halos have significantly smaller $V_{\rm cir, DM}(2~{\rm kpc})$. The $2\sigma$ scatter in $V_{\rm cir, DM}(2~{\rm kpc})$ (from the scatter in the concentration-mass relation) is about a factor of $1.8$ in SIDM similar to that in CDM (about $1.6$).  

Observationally, $V_{\rm cir}(2~{\rm kpc})$ (total rotation velocity) spans from $20$ to $70~{\rm km/s}$ for $\vmax\sim70~{\rm km/s}$~\cite{Oman:2015xda}. The SIDM prediction for the lowest $V_{\rm cir}(2~{\rm kpc})$ is consistent with 20 km/s. When baryons are included, the upper end of the predicted range for $V_{\rm cir}(2~{\rm kpc})$ changes significantly.  
Beyond contributing directly to the total $V_{\rm cir}$, its presence changes the total potential $\Phi_{\rm tot}$ and the equilibrium isothermal solution is consequently denser~\cite{Kaplinghat:2013xca}. The net effect is to increase the upper end of the predicted range for $V_{\rm cir}(2~{\rm kpc})$ to $70~{\rm km/s}$, fully consistent with the data.

For larger galaxies, even the low end of the predicted $V_{\rm cir, DM}(2~{\rm kpc})$ can be changed by the presence of the baryons. We illustrate this in the bottom panel of Fig.~\ref{fig:sidmgeneric}. We adopt $\vmax=120~{\rm km/s}$ and median concentration for the halo. We set the total disk mass to be $10^{10}M_\odot$ (typical for this halo mass) and vary the scale radius of the thin disk $R_d=2$, $3$, and $6~{\rm kpc}$. We use the matching procedure described in Sec.~II to obtain the SIDM halo mass profiles. 
With $R_d=2~{\rm kpc}$, the SIDM prediction for $V_{\rm cir, DM}(2~{\rm kpc})$ is very close to the CDM prediction. The scatter in $V_{\rm cir, DM}(2~{\rm kpc})$ from changing $R_d$ is almost a factor of $2$; even the disk with $R_d=6~{\rm kpc}$ has some effect on the SIDM halo. Thus, the SIDM inner halo mass profile is strongly correlated with the distribution of the baryons, which, along with the scatter from the concentration-mass relation, leads to the diversity in the SIDM halo properties. An upcoming work~\cite{creaseyetal} will explore these facets of thermalization due to self-interactions in N-body simulations of galaxies. 

\begin{figure*}[thb]
\centering
\begin{tabular}{c}
\includegraphics[scale=1.1]{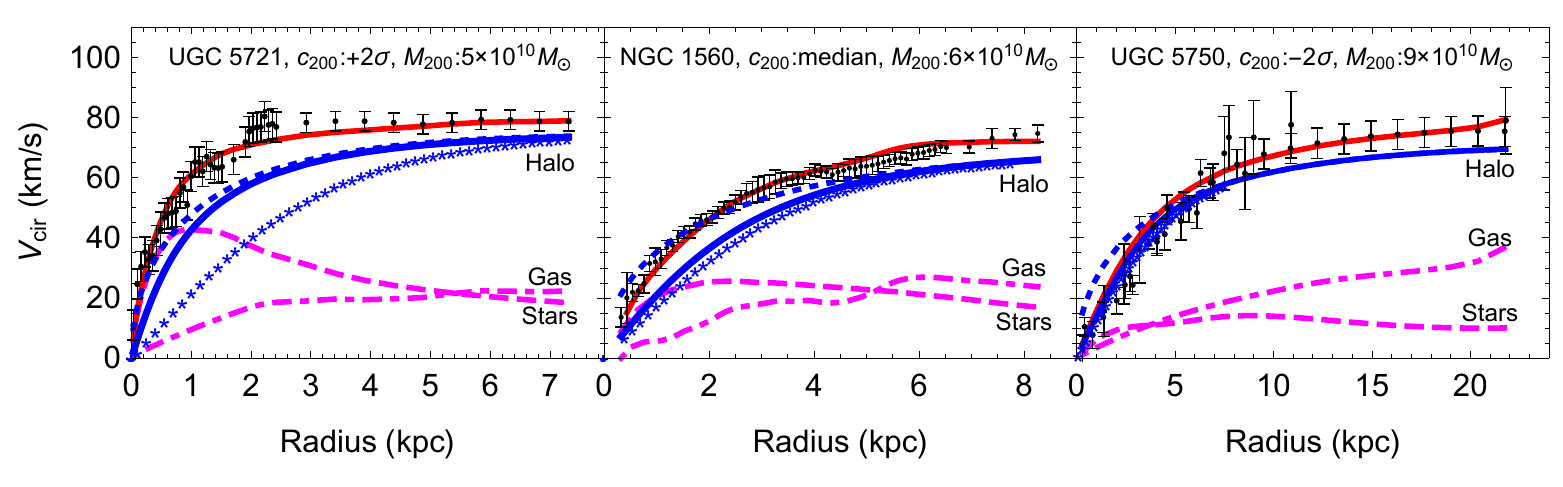}
\end{tabular}
\caption{\label{fig:examples} SIDM fits (red solid) to the rotation curves of spiral galaxies UGC 5721, NGC 1560, and UGC 5750, all with $V_f \approx 80~{\rm km/s}$ but showing extreme diversity in the inner parts. Line types are the same as Fig.~\ref{fig:examplesv130}. 
}
\end{figure*}

\stepcounter{sec}
{\bf \Roman{sec}. Solving the diversity problem in SIDM models.\;}To explicitly demonstrate how the diversity is accommodated in SIDM, we fit to the rotation curves of $30$ galaxies that maximize the diversity and have $V_f$ in the $25\textup{--}300$ km/s range. 
We obtained excellent fits overall, with $\chi^2/{\rm dof}<1$ for $23$ galaxies (DDO 52, 154, 87 126; UGC 128, 5005, 11707, 4483, 3371, 5721, 12506, 1281; UGCA 442; NGC 2366, 7331, 2403, 3109, 1560, 2903, 3198; F583-1, F579-V1, M33), and $\chi^2/{\rm dof}<2$ for the rest (UGC 2841, 5750; NGC 6503, F571-8, F563-V2, DDO 133, IC 2574). 
In Figs.~\ref{fig:examplesv130} and \ref{fig:examples}, we show the fits to some of the most extreme examples highlighted in~\cite{Oman:2015xda}\footnote{In the Appendix, we  show the fits for the $24$ other galaxies.}. 
For each galaxy, we compute the thin disk parameters ($\Sigma_0, R_d$) that best match the rotation curves of the stellar disk in the literature, given a value of the mass-to-light ratio ($\Upsilon_*$). In computing $\rho_{\rm iso}$, we have neglected the potential of the gaseous disk and stellar bulge, which is a good approximation for the fits shown here. In our fits, the outer halo $V_{\rm max}$ is essentially set by the measured $V_f$ and the freedom in the fits is primarily due to $\Upsilon_*$ and the scatter allowed in the concentration of the outer halo.

NGC 6503~\cite{begeman1987} and UGC 128~\cite{vanderhulst1993} clearly illustrate the diverse features in the rotation curves caused by the baryon distribution in Fig.~\ref{fig:examplesv130}. Both galaxies have $V_f\approx 130~{\rm km/s}$, but their inner rotation curves are very different. For NGC 6503, the circular velocity increases sharply in the inner regions and reaches its asymptotic value around $5~{\rm kpc}$; in UGC 128, it increases very mildly and reaches $V_f$ at $20~{\rm kpc}$. Despite the dramatic differences, the SIDM halo with {\em median} concentration provides a remarkable fit to both galaxies. NGC 6503 is a high surface brightness galaxy and its inner gravitational potential is dominated by the stellar disk, which contributes significantly to the observed $V_{\rm cir}$. Moreover, the inner SIDM (isothermal) halo density in the presence of the disk is almost an order of magnitude larger than when neglecting the influence of the disk, which boosts the halo contribution at $V_{\rm cir}(2~{\rm kpc})$ from $20$ to $60~{\rm km/s}$. 
In contrast, the stellar disk has a negligible effect on the SIDM halo of UGC 128. 

It is interesting to note that in NGC 6503 the rotation curve becomes flat at $r\approx 3~{\rm kpc}$, which implies that the total density profile scales as a power-law in radius, with index close to $-2$, from inner regions dominated by the disk to outer regions dominated by dark matter. Thus, the thermalization of dark matter in SIDM models provides a natural mechanism for understanding the long-standing puzzle of the disk-halo conspiracy~\cite{1986RSPTA.320..447V}.  
This power-law behavior of the total mass density is prevalent in large spiral and elliptical galaxies~\cite{Humphrey:2009te,2015ApJ...804L..21C}. We show the SIDM fit to the rotation curve of massive spiral galaxy NGC 2903 \cite{deblok2008} in right panel of Fig.~\ref{fig:examplesv130} as an example. 

In Fig.~\ref{fig:examples}, we show SIDM fits for UGC 5721 \cite{swaters2009_whisp, swaters2003}, NGC 1560 \cite{gentile2010}, and UGC 5750 \cite{vanderhulst1993, mcgaugh2001, deBlok:2002tg, kuziodenaray2006}. All have similar $V_f\approx 80~{\rm km/s}$, but the shapes of the rotation curves are very different in the inner regions. UGC 5721 and UGC 5750 are at opposite extremes for the rotation curve diversity in this mass range. Despite the diversity, the SIDM halo model provides an impressive fit to the rotation curves. We find that NGC 1560 has a median halo, UGC 5721 has a denser halo, and UGC 5750 has an underdense halo, but all within $2\sigma$ of the median expectation. The observed $\vc(2~{\kpc})$ is close to $20~{\rm km/s}$ for UGC 5750, while the corresponding CDM halo has $\vc(2~{\kpc})\approx 30~{\rm km/s}$ even with a  concentration $2\sigma$ lower than the median value. 
The effect of the disk is most significant in UGC 5721, resulting in a SIDM halo {\em similar} to the CDM one and a flat $V_{\rm cir}$ even at $2~{\rm kpc}$. The effect becomes mild in NGC 1560 and negligible in UGC 5750, consistent with their luminosities. We have further checked that UGC 5721 can also be fit with a $1.5\sigma$ higher $c_{200}$ value and $M_{200}=6\times10^{10}\msun$, and UGC 5720 with a $1.5\sigma$ lower $c_{200}$ value and $M_{200}=8\times10^{10}\msun$. This is due to a mild $c_{200}\textup{--}M_{200}$ degeneracy.

\begin{figure}[thb]
\centering
\begin{tabular}{c}
\includegraphics[scale=0.65]{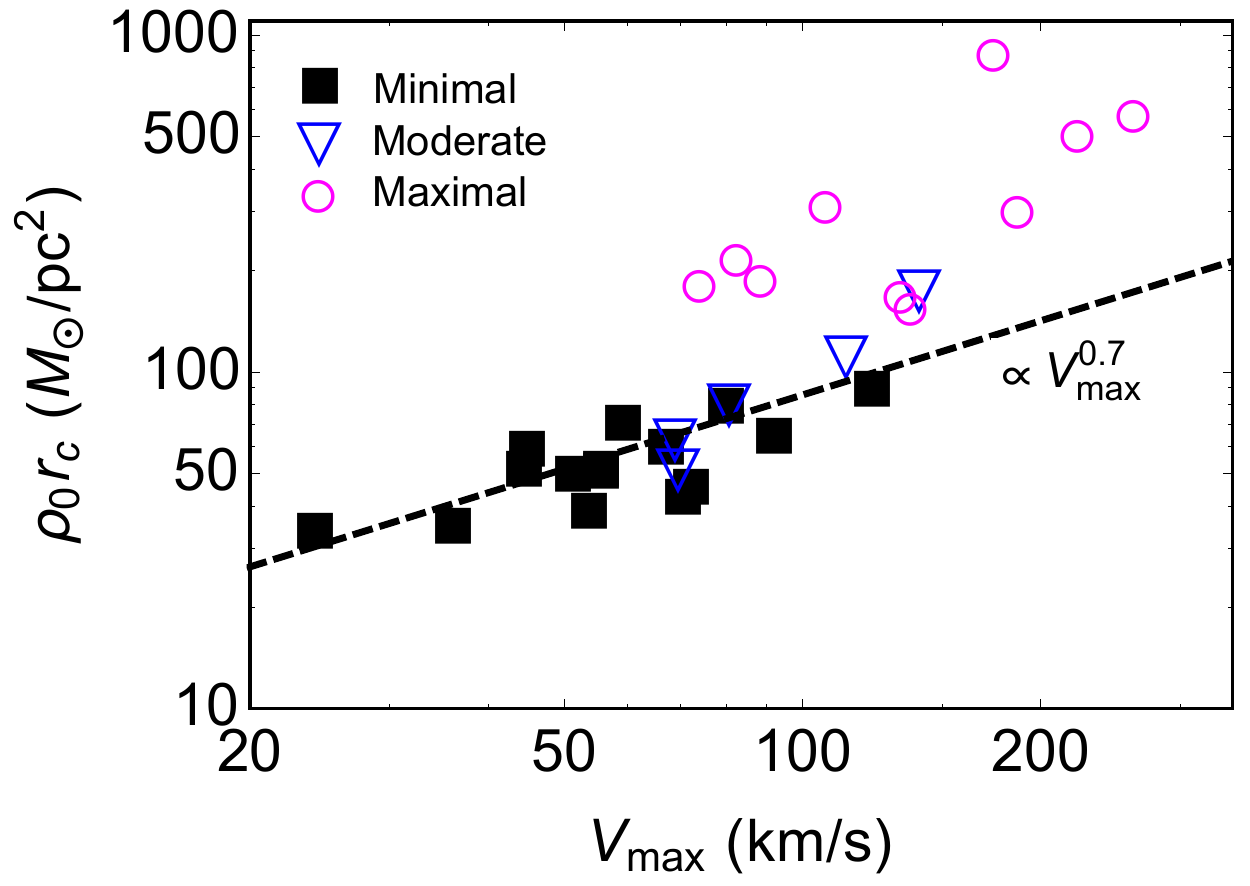}
\end{tabular}
\caption{\label{fig:uniformity} The inferred SIDM core density times core radius (``surface density") for the $30$ galaxies we have fit. The ``minimal" sample is composed of dark matter dominated galaxies for which baryons don't change the SIDM profile significantly. The surface densities of these galaxies scale as $V_{\rm max}^{0.7}$ (dashed line), which can be traced to the concentration-mass relation. The ``intermediate" (triangles) and ``maximal" (circles) samples show progressively increasing effects of the stellar disk on the SIDM halo.
}
\end{figure}

\stepcounter{sec} 
{\bf \Roman{sec}. Diversity from uniformity.\;}The diversity problem is solved by a combination of features in $\Lambda$SIDM that are not separate pieces to be tuned but instead arise from the requirement that the inner halo at $r \lesssim r_1$ is thermalized. While the inner rotation curves display great diversity for the same halo mass, there are also remarkable similarities. In Fig.~\ref{fig:uniformity}, we plot a measure of the surface density of dark matter defined as $\sdm = \rho_0 r_c$, where $\rho_0$ is the central density inferred from the fits and $r_c$ is the core radius where the dark matter density is half of $\rho_0$. We split the $30$ galaxies into three samples by the effect that the baryons have on the halo profiles: the ``minimal" sample is where the halo profiles are essentially unchanged by the baryons, while the ``maximal" sample shows a roughly $1/r^2$ scaling for the total density profile over most of the halo. The changes to the dark matter density profile of the ``moderate" sample is in between that of the minimal and maximal samples.

The minimal sample shows a clear scaling relation for $\sdm$ vs. $V_{\rm max}$ (of the NFW halo), which is a reflection of the concentration-mass relation~\cite{Lin:2015fza}. Our model predicts $\sdm \propto V_{\rm max}^{0.7}$, which may be roughly understood from the approximate scalings $r_c \propto r_s$~\cite{Rocha:2012jg} and $\rho_0 \propto V_{\rm max}^2/r_s^2$ from dimensional arguments. However, there is clear deviation when baryons become important, since $\rho_0$ increases and the core radius is set by the gravitational potential of the baryons. Our $\sdm$ values for the minimal sample are consistent with previous results~\cite{Donato:2009ab}.

\stepcounter{sec} 
{\bf \Roman{sec}. Self-interaction cross section.\;} We fixed $\sigma/m=3~{\rm cm^2/g}$ in our analysis and it provided good fits for all $30$ galaxies with $c_{200}$ values within the $2\sigma$ range, and mass-to-light ratios in the range preferred by recent measurements \cite{Martinsson:2013ukc,2014ApJ...788..144M,2014AJ....148...77M}. 
Galaxies with low $V_{\rm cir}(2~{\rm kpc})$ like UGC 5750 and IC 2574 drive the preference for this large $\sigma/m$.
However, there are degeneracies among $\sigma/m$, $\Upsilon_*$ and $c_{200}$. For higher luminosity galaxies, in particular those with $V_f\gtrsim 200~{\rm km/s}$ such as NGC 2903, 7331, 2841, and UGC 12560, good fits can also be found with smaller cross sections, $\sigma/m\sim1~{\rm cm^2/g}$, by varying $\Upsilon_*$ very mildly. This implies that a mild velocity dependence, which would be required by the galaxy cluster constraints in specific models~\cite{Kaplinghat:2015aga,Boddy:2016bbu}, is also consistent with the data. We have checked that the $\Upsilon_*$ values (disk masses) required by the fits are consistent with abundance matching expectations~\cite{Garrison-Kimmel:2013eoa}. We leave a detailed exploration of the cross section, abundance matching and cosmological dependence for a future statistical analysis of a larger sample of galaxies. 

\stepcounter{sec}
{\bf \Roman{sec}. Conclusions.\;}The rotation curves of spiral galaxies exhibit considerable diversity, which lacks an explanation. The problem is most severe for flat rotation velocities in the range $60\textup{--}100~{\rm km/s}$, with almost an order of magnitude spread in the inferred dark matter core densities. To address this problem in the context of SIDM models, we developed numerical templates for modeling the SIDM halo including the presence of a stellar disk and fit a wide variety of rotation curves for spiral galaxies that exemplify the diversity over three orders of magnitude in total mass. 
Our model utilizes the $\Lambda$CDM concentration-mass relation and a fixed self-interaction cross section. We have demonstrated that the variation in the distribution of baryons and the reaction of the SIDM halo to it, when melded with the expected scatter in the concentration-mass relation due to assembly history of halos, can explain the diverse dark matter distributions in spiral galaxies. 

{\it Acknowledgments}:  We thank Erwin de Blok, Gianfranco Gentile, Federico Lelli, Stacy McGaugh, Se-Heon Oh, and Rob Swaters for providing us the rotation curve data.  This work was supported by IBS under the project code IBS-R018-D1 (AK), the National Science Foundation Grant PHY-1620638 (MK) and PHY-1522717 (ABP), and the U. S. Department of Energy under Grant No.~DE-SC0008541 (HBY).  ABP acknowledges support from a GAANN fellowship and the Mitchell Institute for Fundamental Physics and Astronomy at Texas A\&M University. HBY acknowledges support from the Hellman Fellows Fund.

\bibliography{diversity_table}

\renewcommand{\figurename}{{\bf Appendix \textup{--}} Figure}
\setcounter{figure}{0}

\begin{figure*}[htb]
\centering
\begin{tabular}{@{}cc@{}}
\includegraphics[scale=0.7]{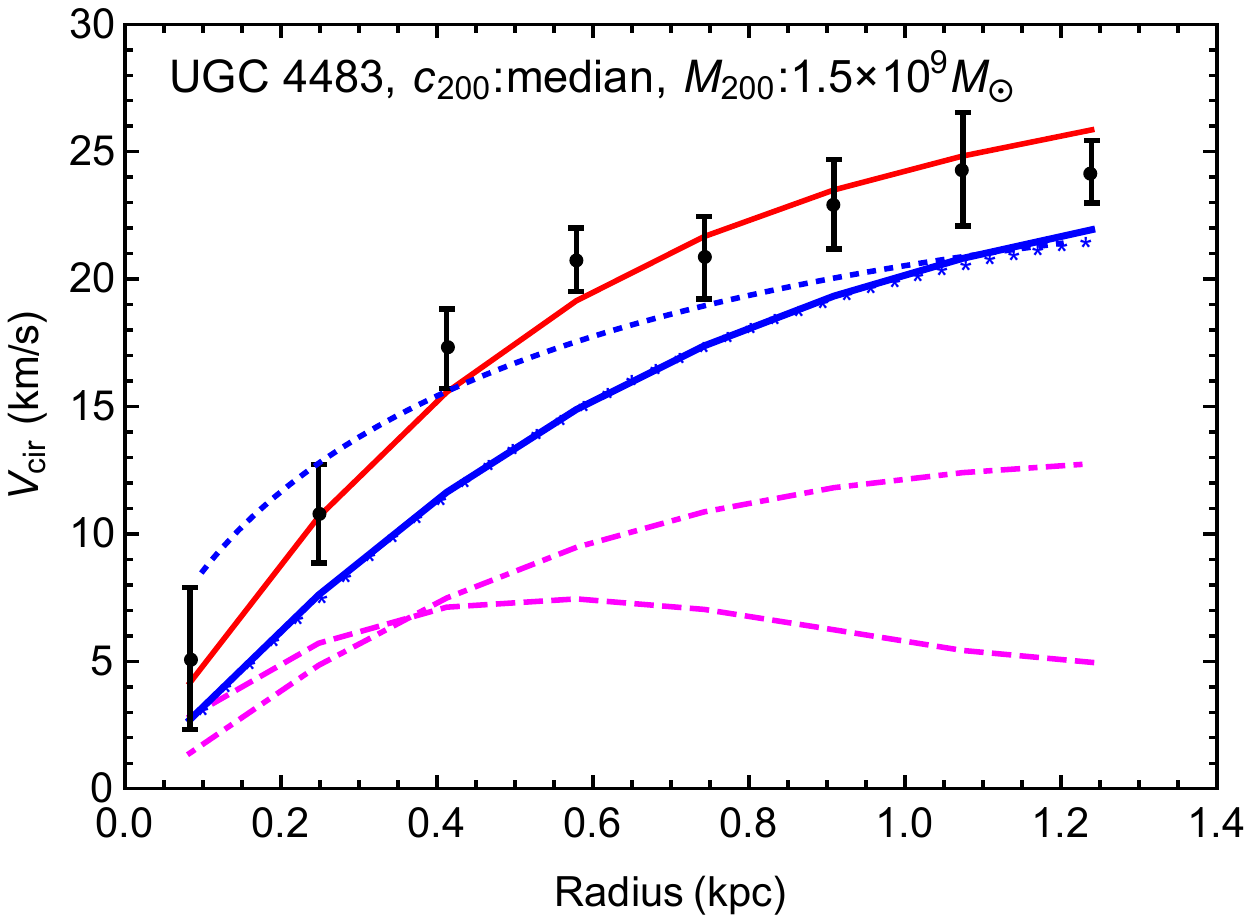}&
\includegraphics[scale=0.68]{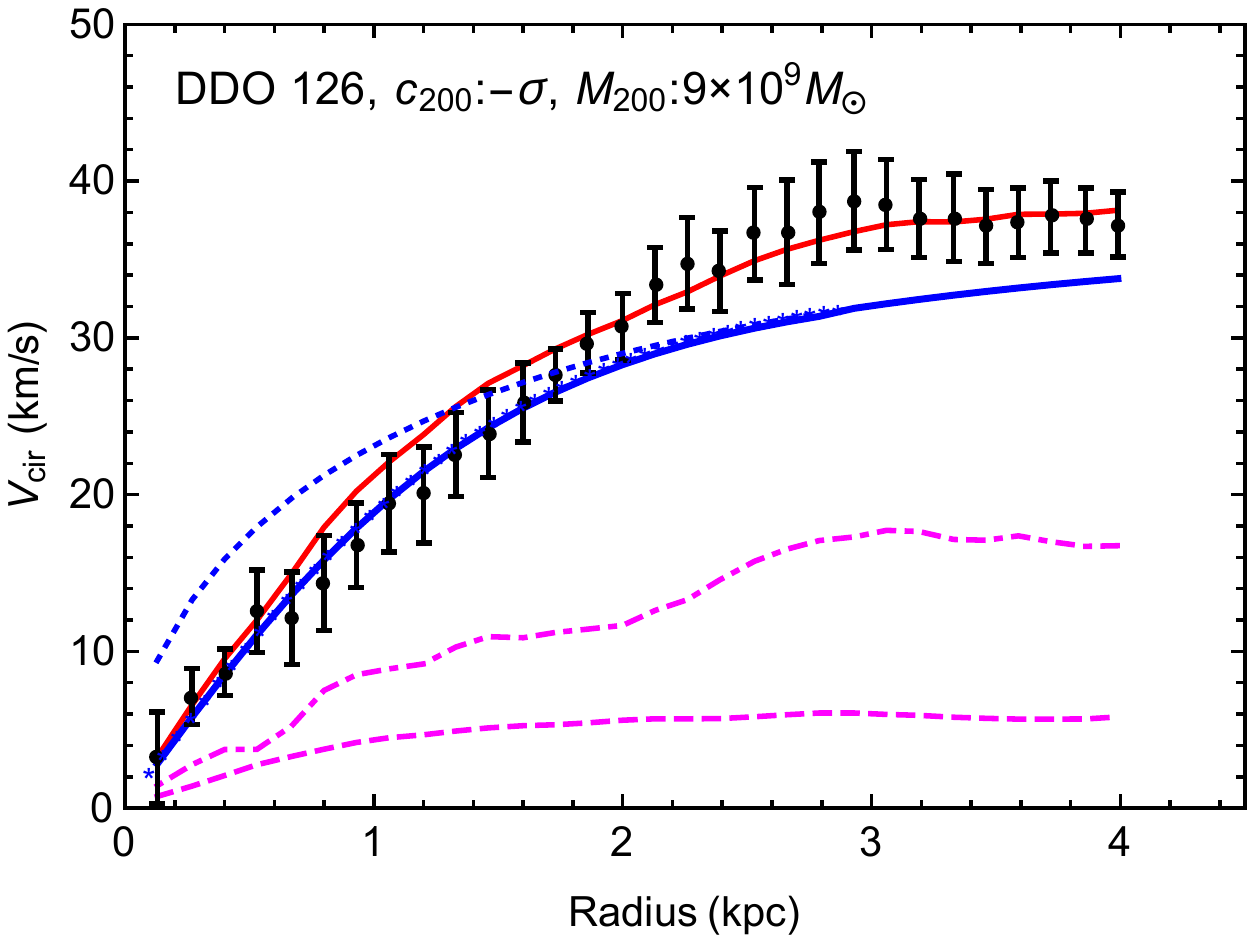}\\
\includegraphics[scale=0.7]{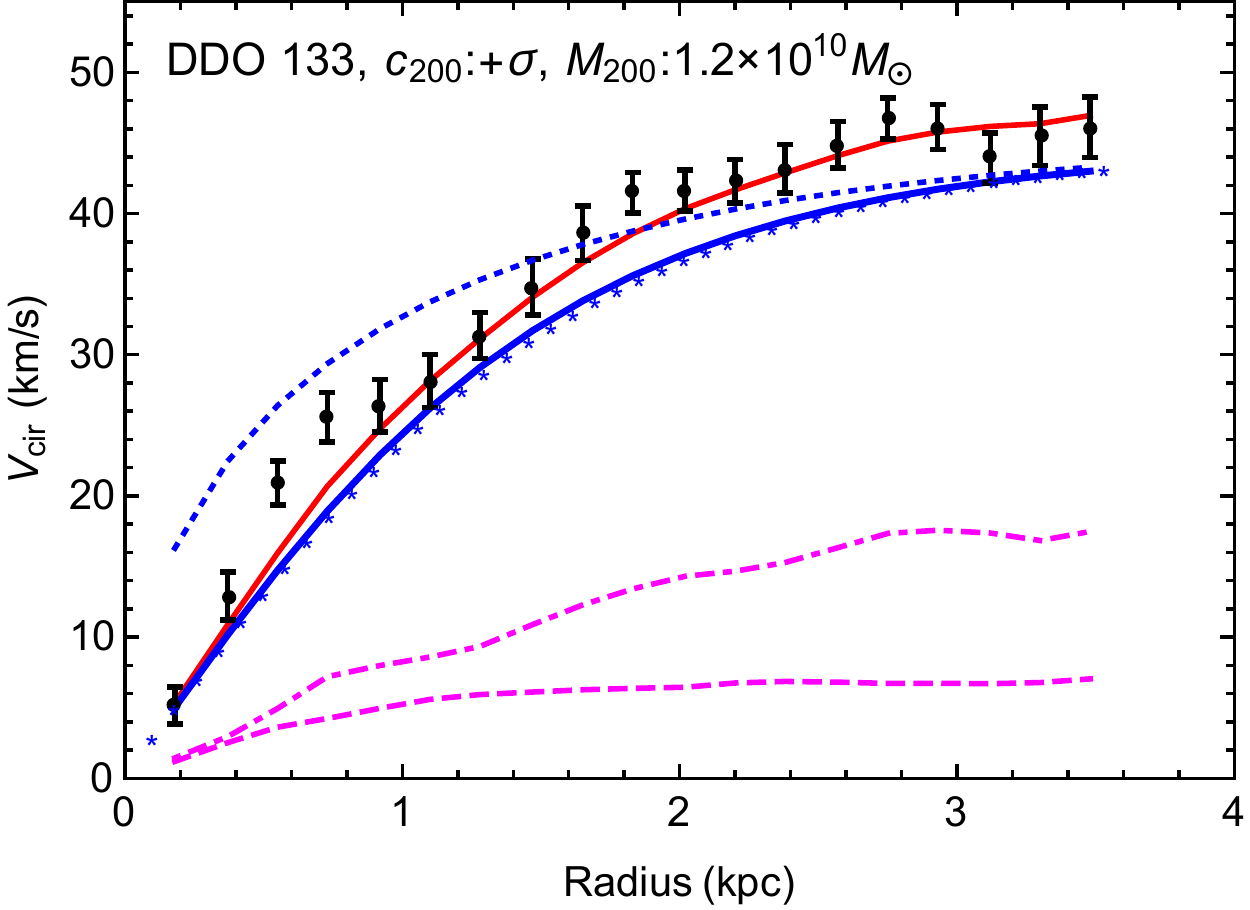}&
\includegraphics[scale=0.7]{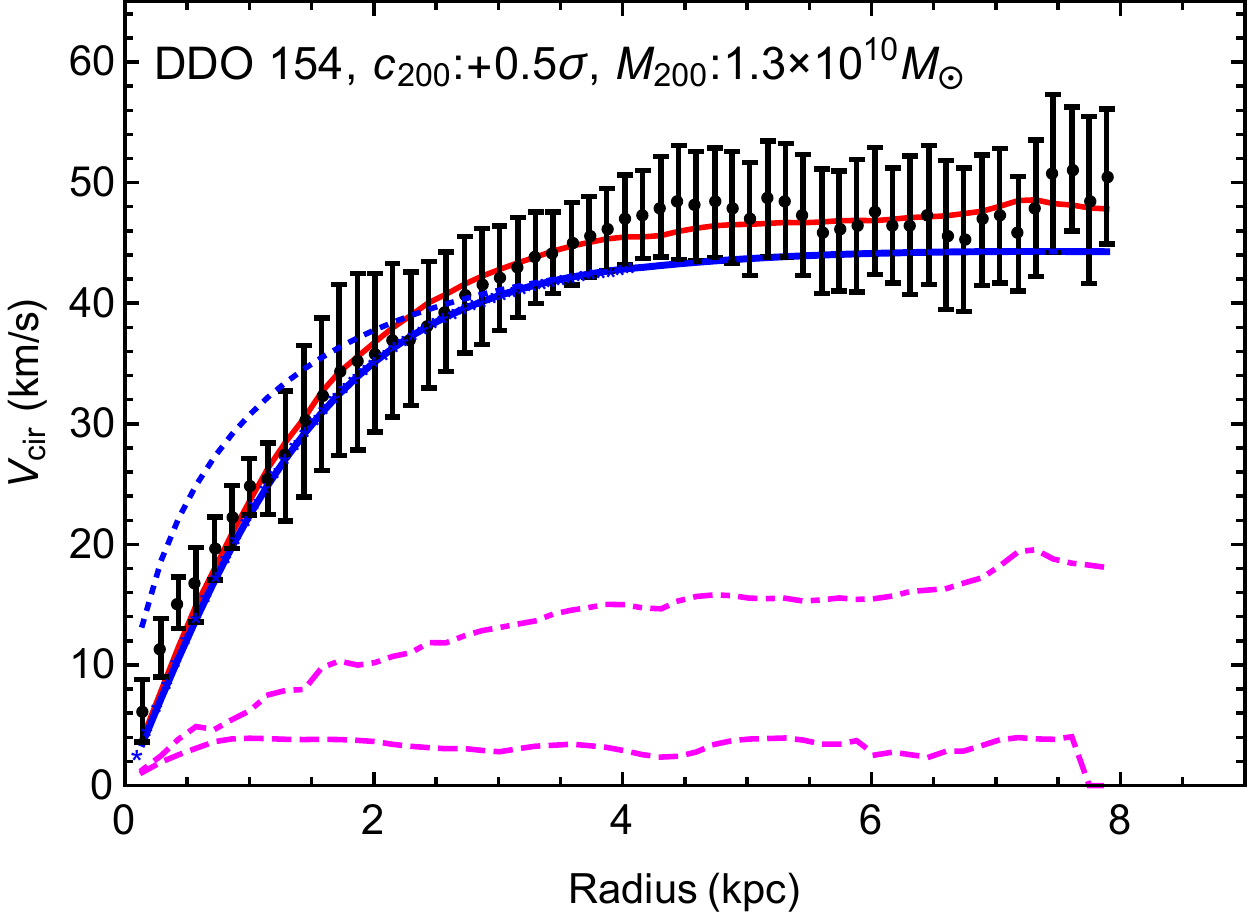}
\end{tabular}
\caption{\label{fig:v25} Galaxies with $V_f\approx25\textup{--}50~{\rm km/s}$. Data for UGC 4483  from \cite{Lelli2012_ugc4483} and data DDO 126, 133, 154 from \cite{oh2015}. The total fit is displayed in red and it includes contributions from the SIDM halo (blue solid), stars (magenta dashed) and gas (magenta dot-dashed). The predictions of the corresponding CDM halo (dotted) and the SIDM halo neglecting the influence of the baryons (asterisk) are also shown.  In our analysis, we do not take into account the influence of the gas component on the isothermal dark matter profile, which could be moderately important for the DDO 126 fit. }
\end{figure*}

\begin{figure*}[htb]
\centering
\begin{tabular}{@{}cc@{}}
\includegraphics[scale=0.7]{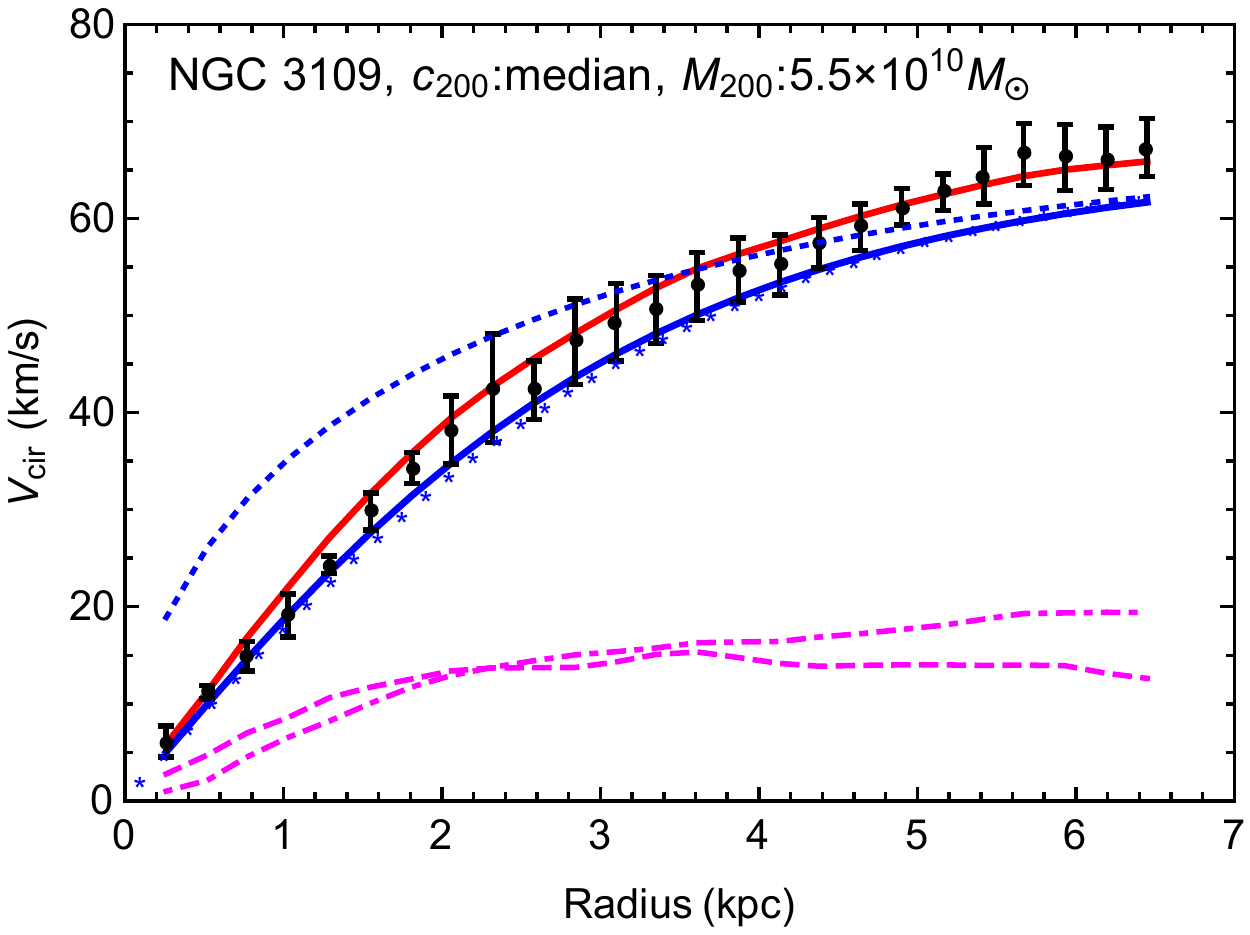}&
\includegraphics[scale=0.7]{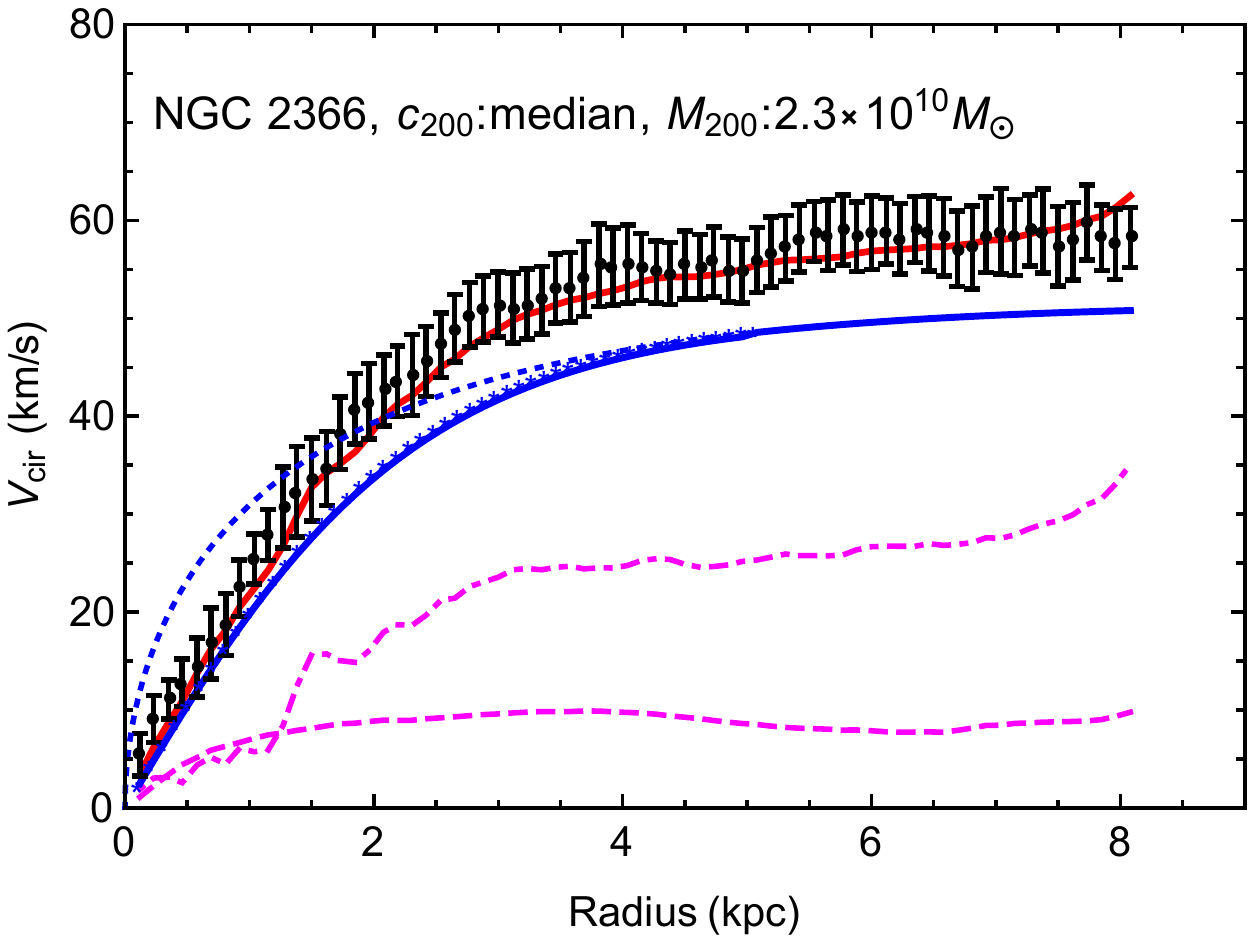} \\
\includegraphics[scale=0.7]{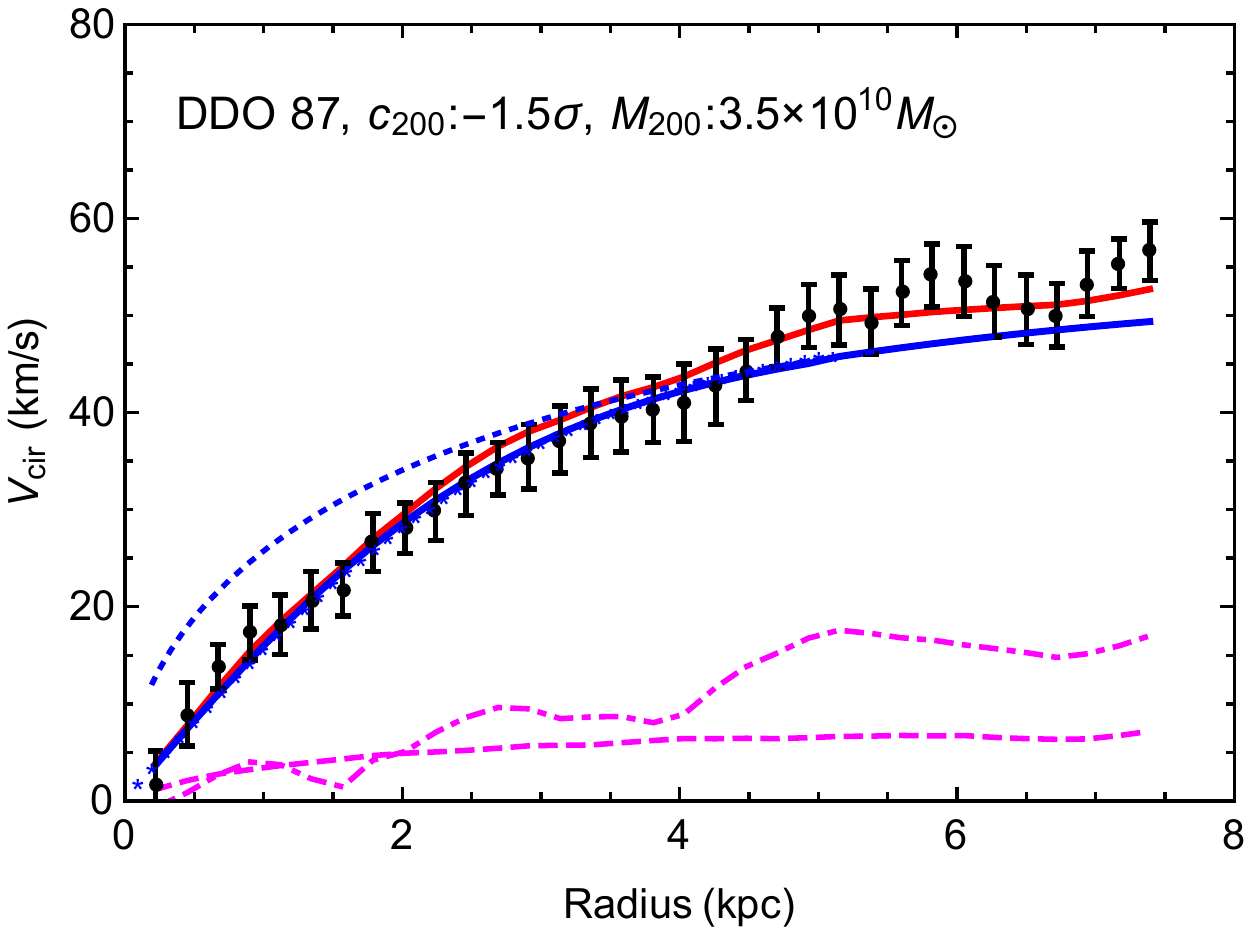}&
\includegraphics[scale=0.7]{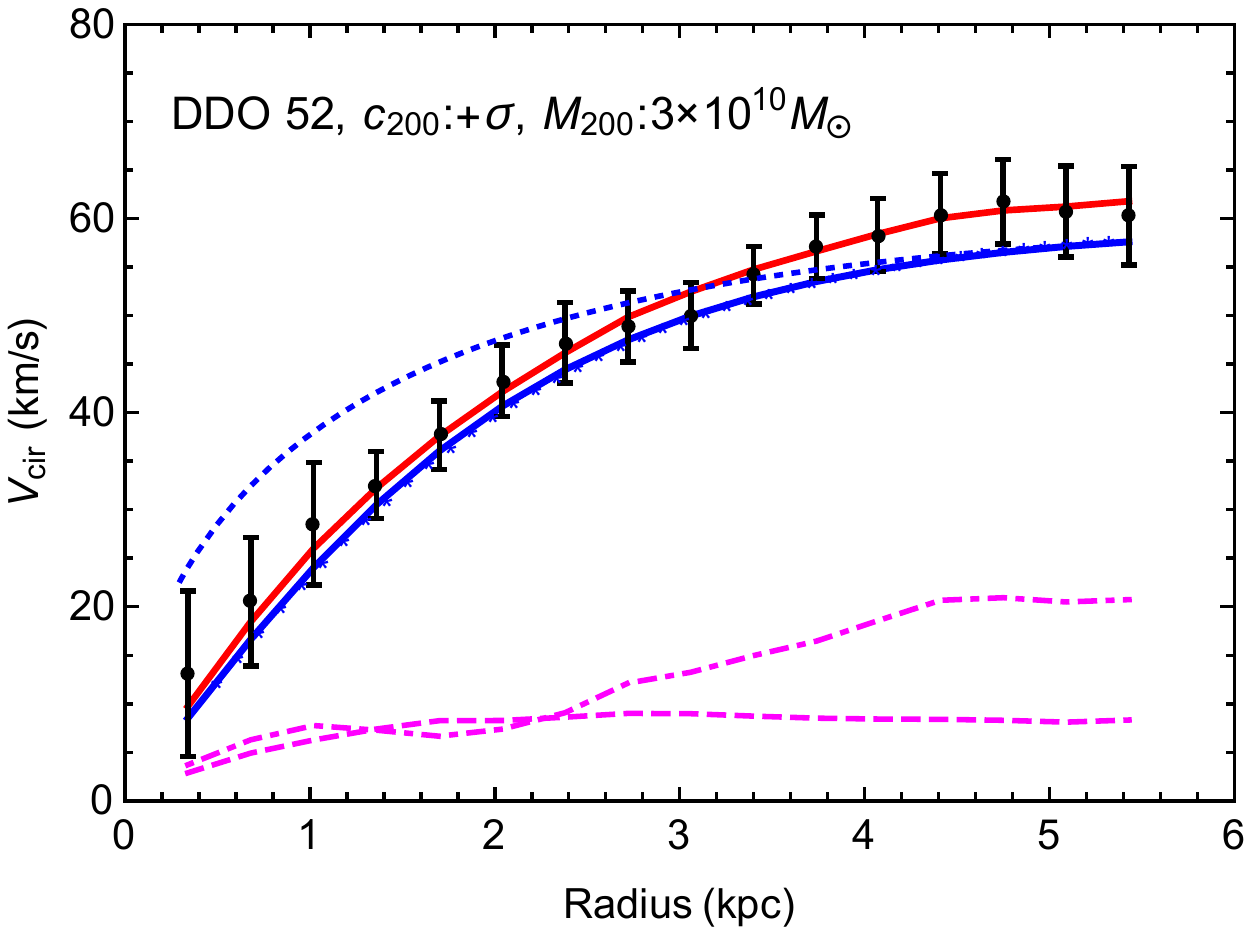} \\
\includegraphics[scale=0.7]{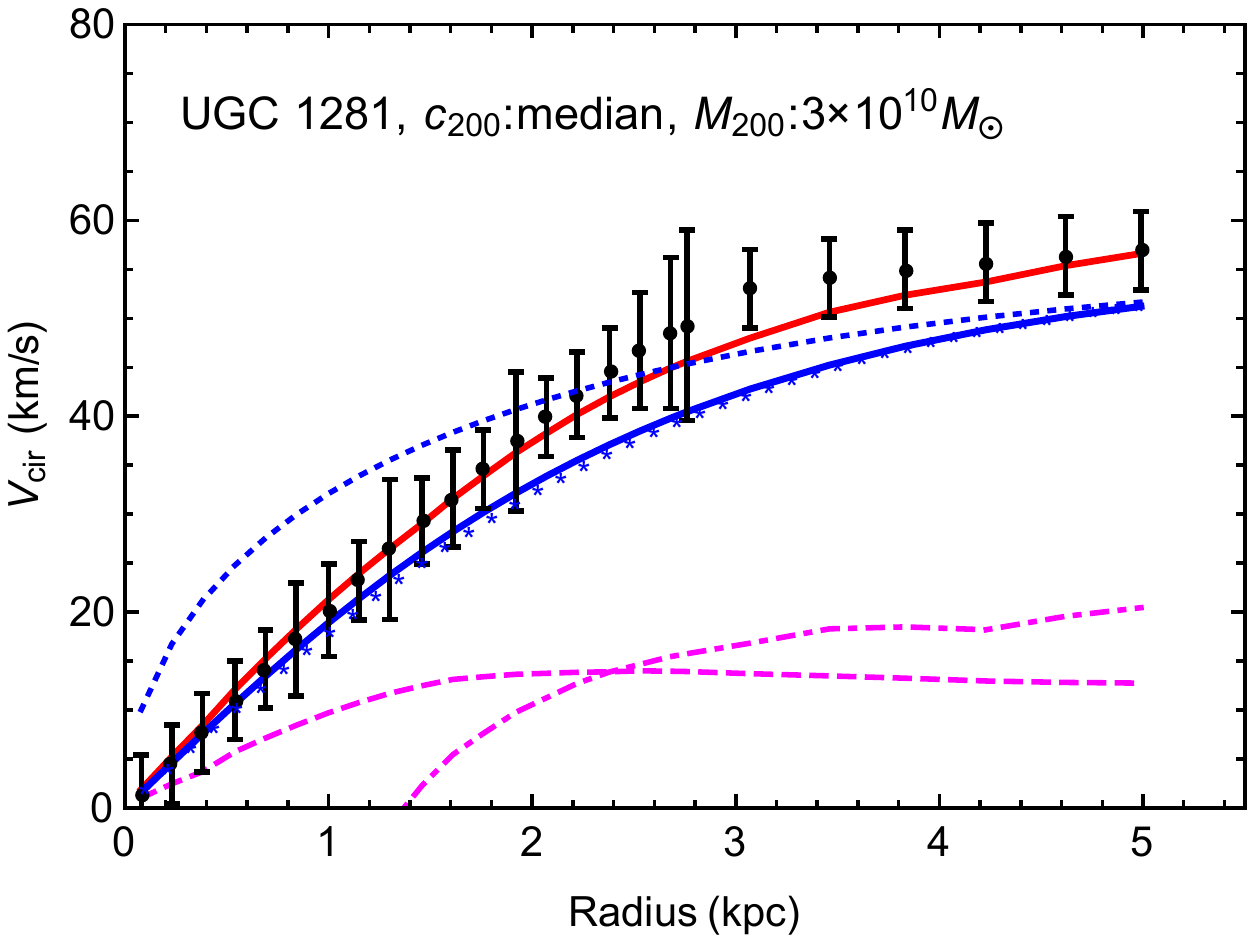} &
\includegraphics[scale=0.7]{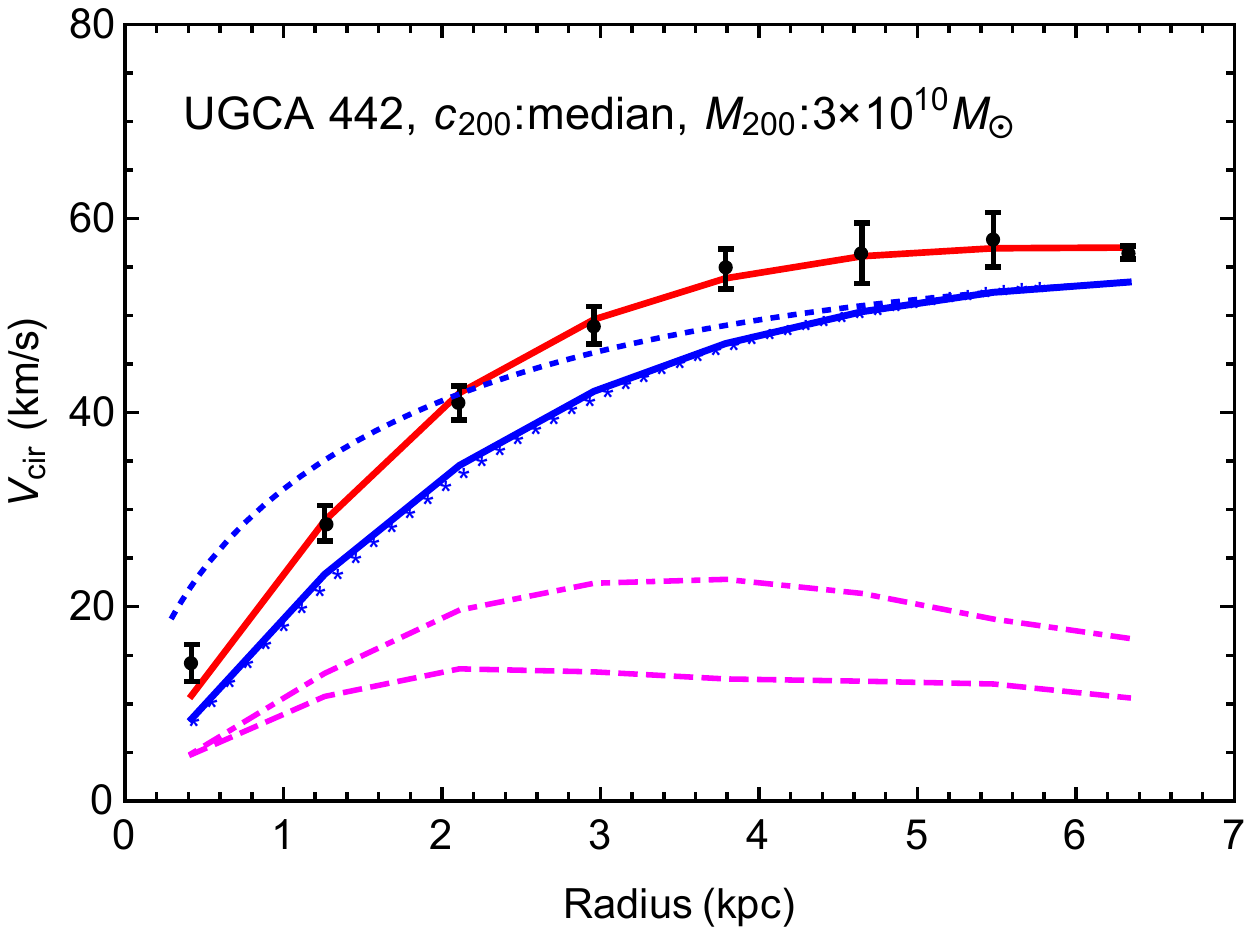} 
\end{tabular}
\caption{\label{fig:v60} Galaxies with $V_f\approx60~{\rm km/s}$. Data for NGC 3109, UGC 1281 and UGCA 442 from \cite{Lelli2016_sparc}, and data for  NGC 2366, DDO 87, and DDO 52 from \cite{oh2015}. The total fit is displayed in red and it includes contributions from the SIDM halo (blue solid), stars (magenta dashed) and gas (magenta dot-dashed). The predictions of the corresponding CDM halo (dotted) and the SIDM halo neglecting the influence of the baryons (asterisk) are also shown. In our analysis, we do not take into account the influence of the gas component on the isothermal dark matter profile, which could be moderately important for the fit to NGC 2366. }
\end{figure*}

\begin{figure*}[htb]
\centering
\begin{tabular}{@{}cc@{}}
\includegraphics[scale=0.55]{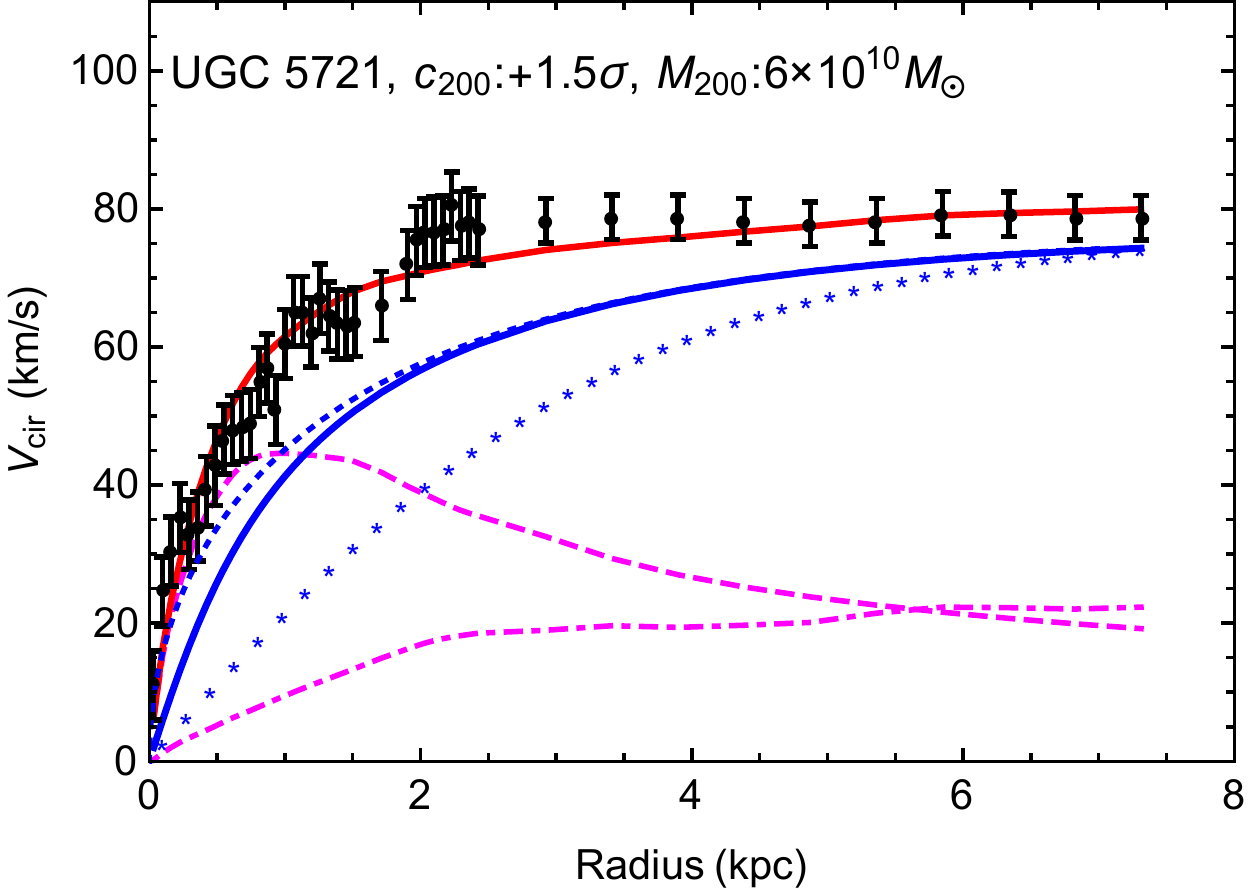}&
\includegraphics[scale=0.55]{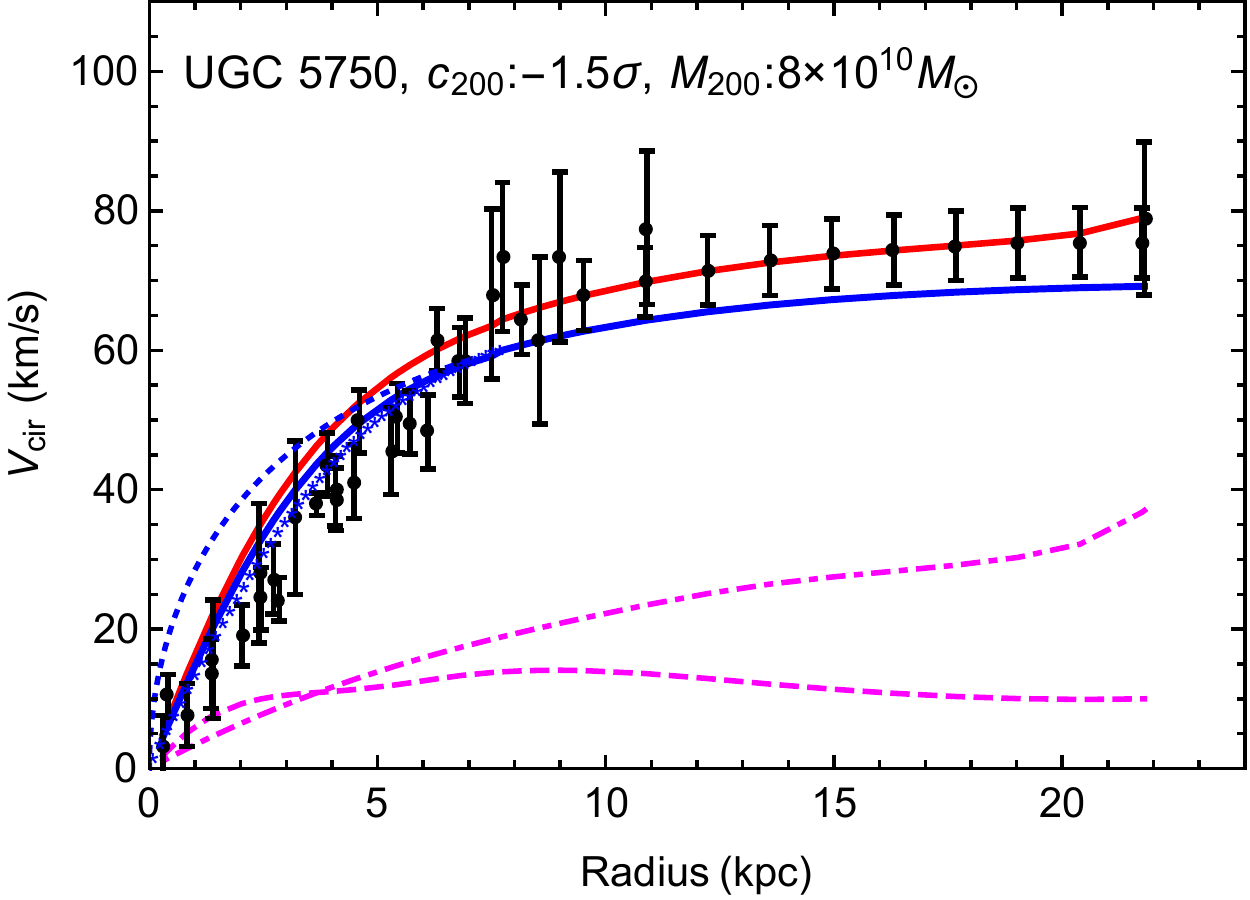} \\
\includegraphics[scale=0.55]{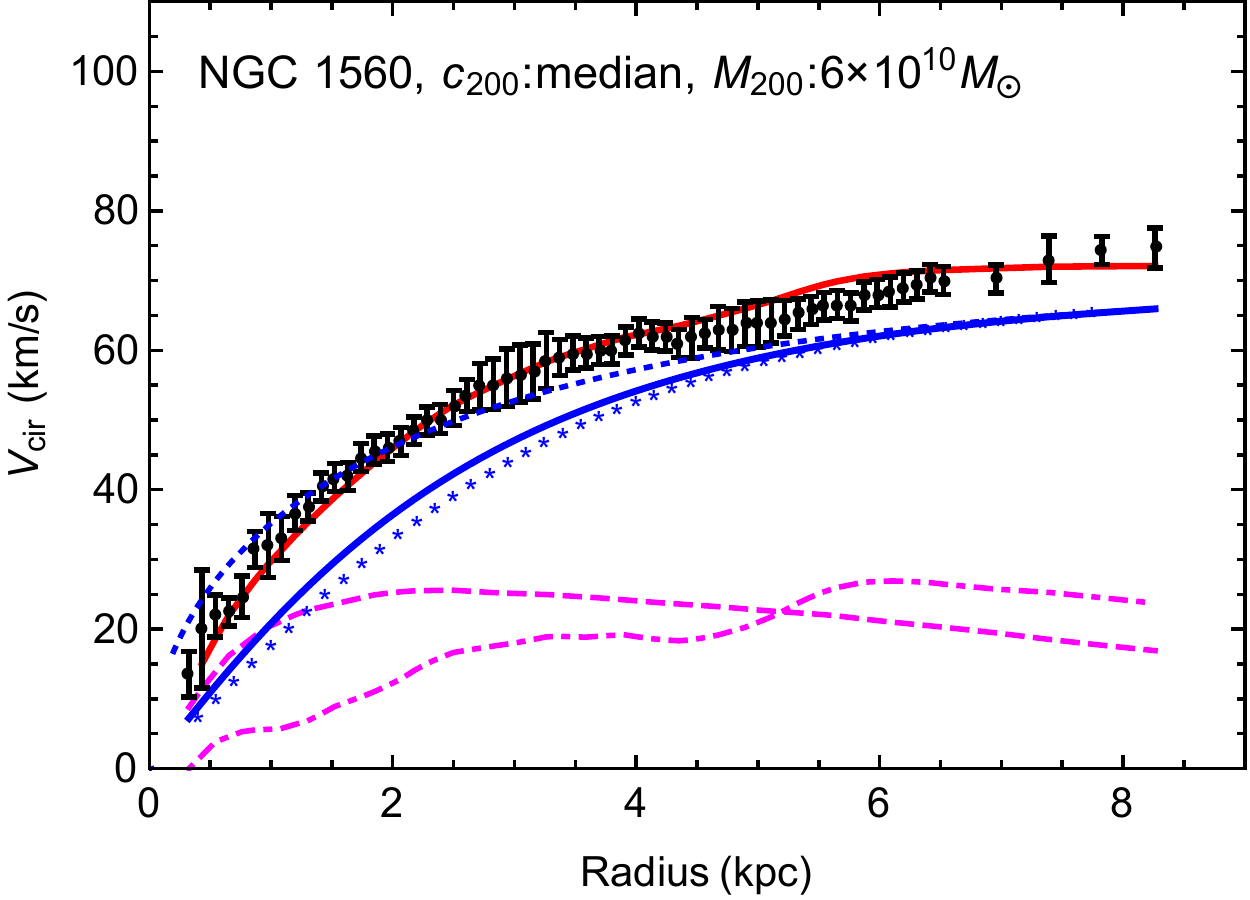}&
\includegraphics[scale=0.55]{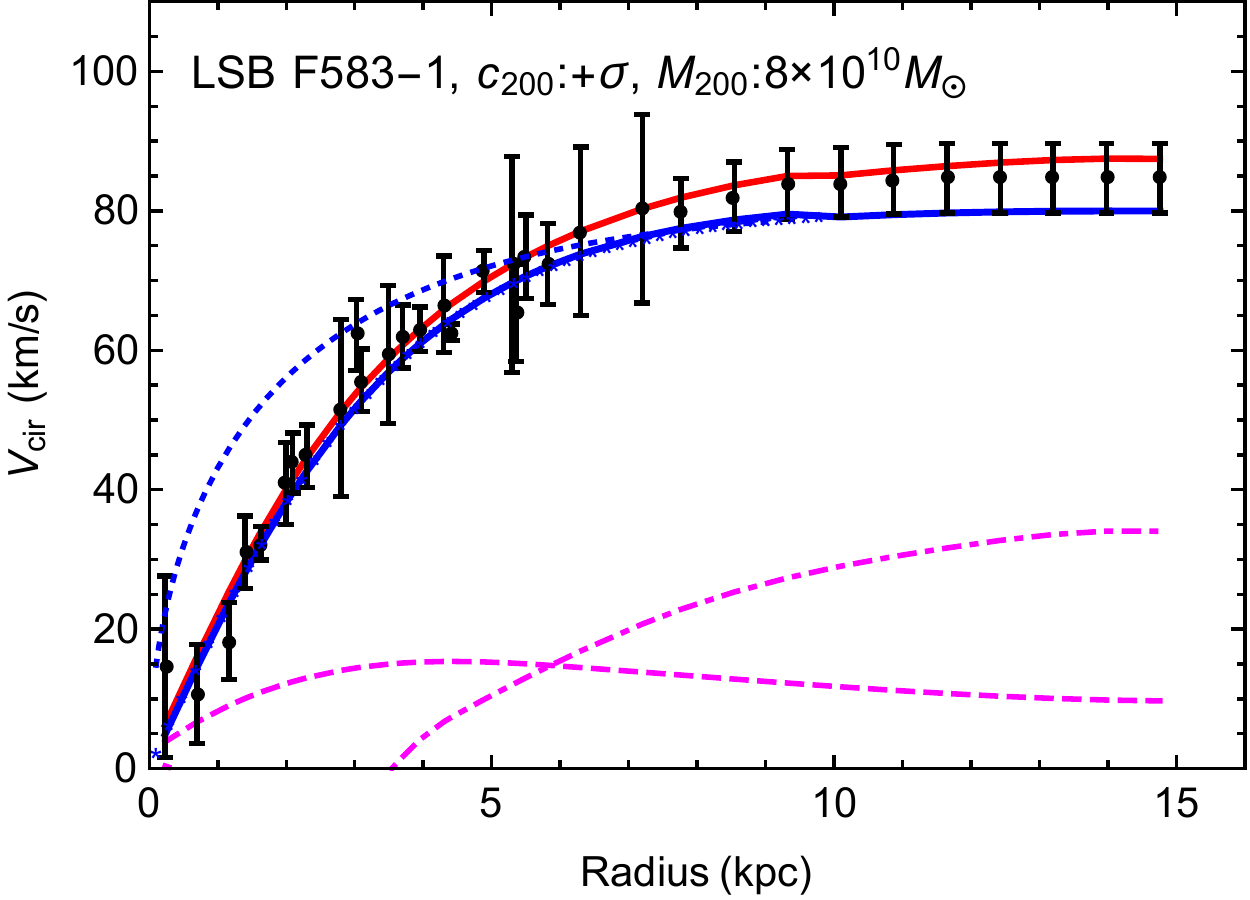} \\
\includegraphics[scale=0.55]{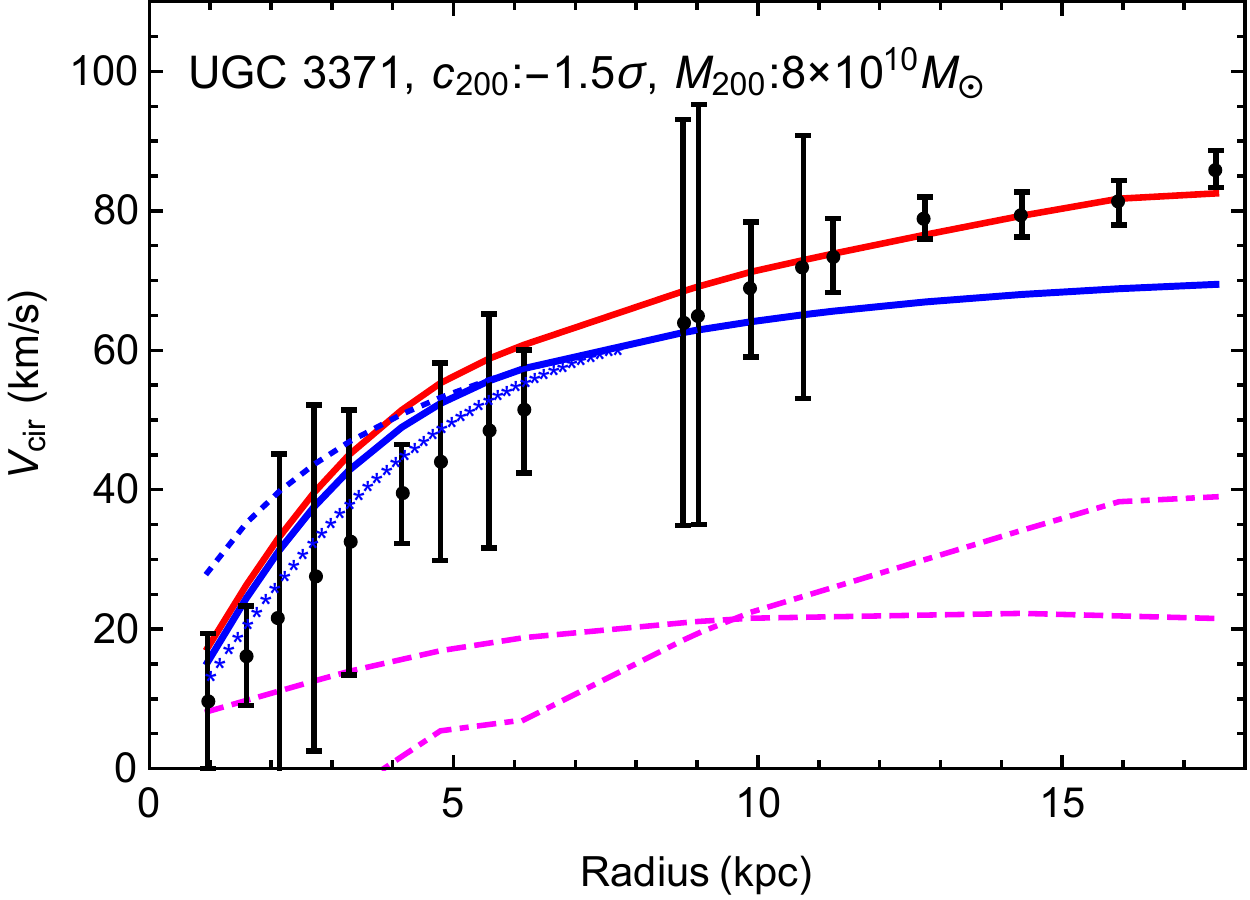} &
\includegraphics[scale=0.55]{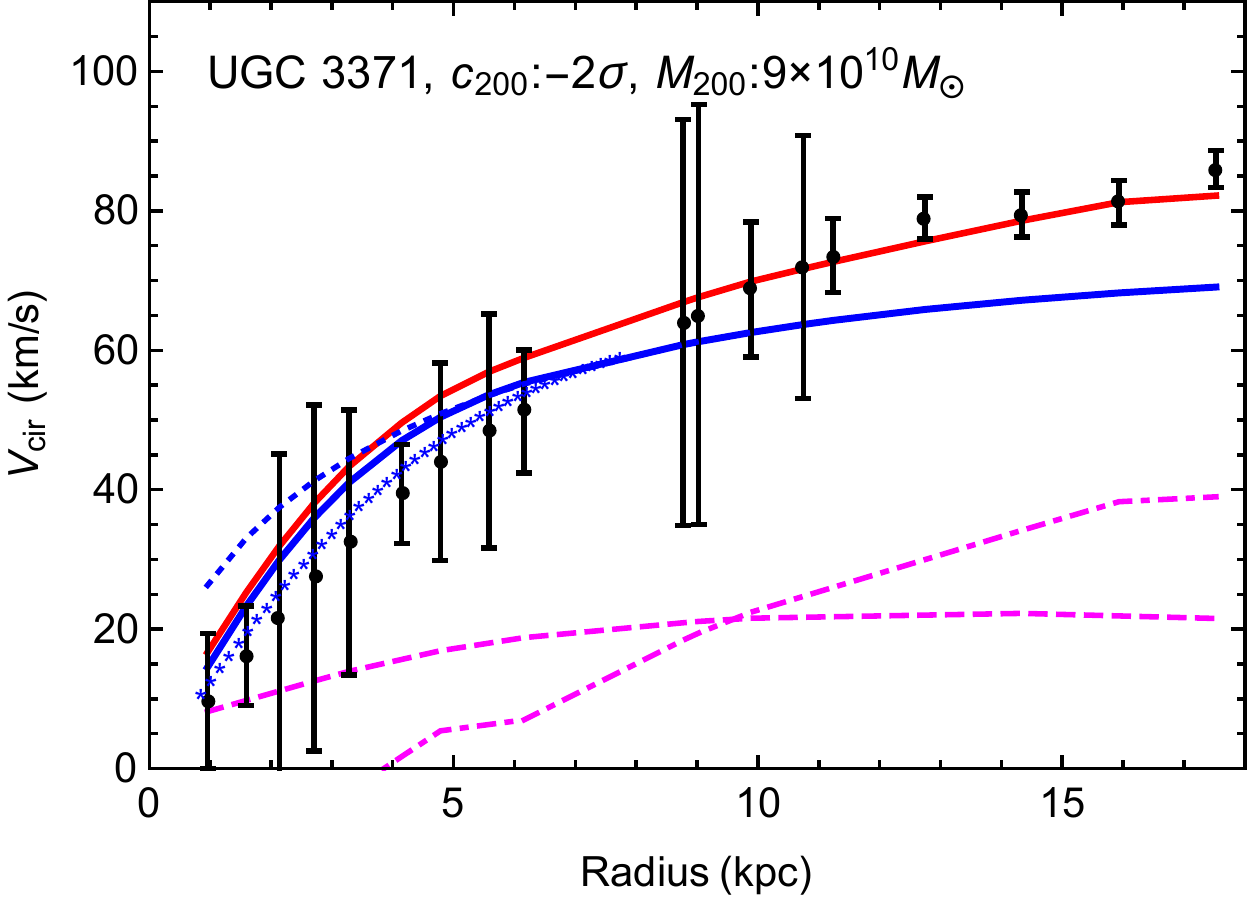}\\
\includegraphics[scale=0.55]{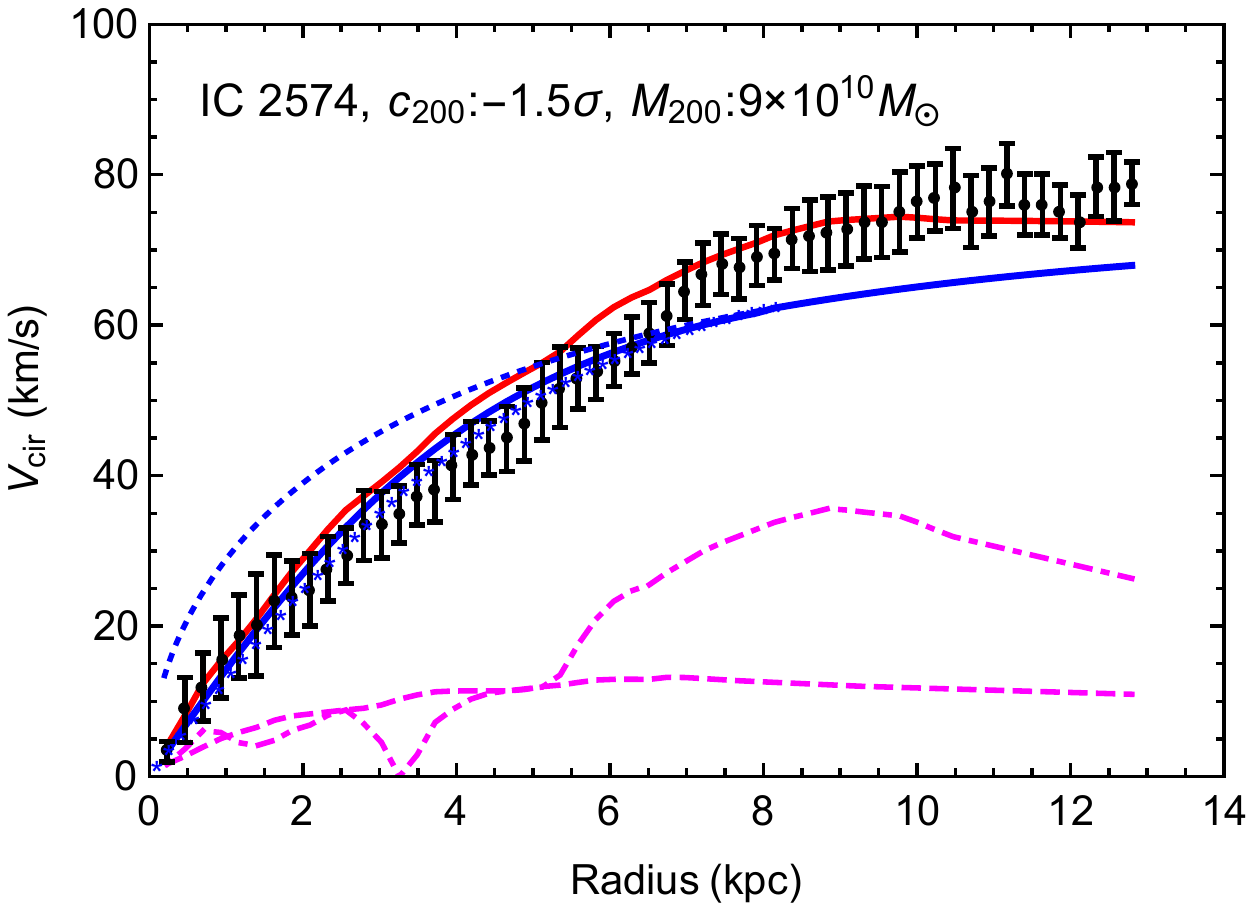}&
\includegraphics[scale=0.55]{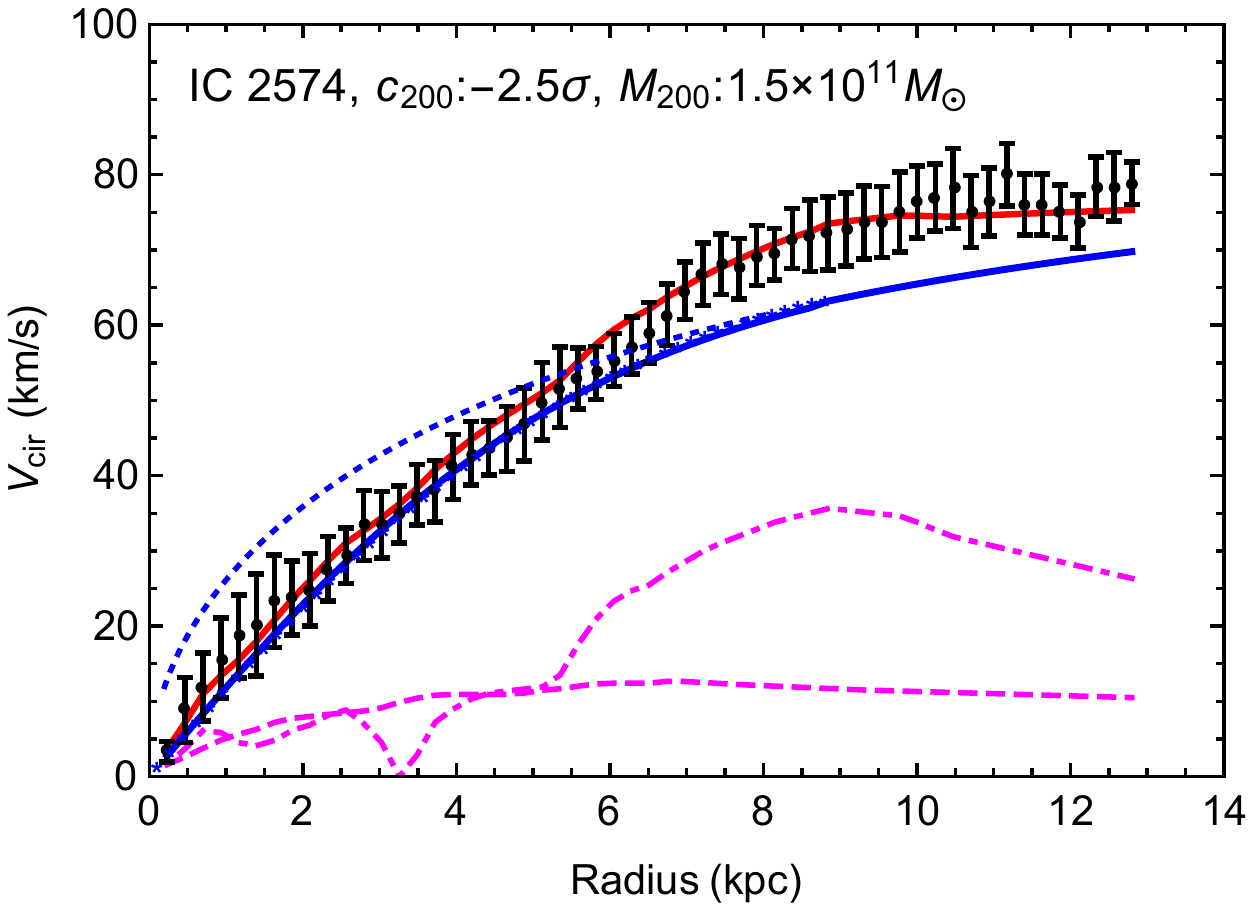}\\
\end{tabular}
\caption{\label{fig:v80} Galaxies with $V_f\approx80~{\rm km/s}$.  Data for UGC 5721, UGC 5750, F583-1, IC 2574, NGC 1560, and UGC 3371 from \cite{swaters2009_whisp, swaters2003}, \cite{vanderhulst1993, mcgaugh2001, deBlok:2002tg, kuziodenaray2006}, \cite{deblok1996, mcgaugh2001, kuziodenaray2006}, \cite{oh2011}, \cite{gentile2010}, and \cite{deBlok:2002tg, swaters2009_whisp} respectively. The total fit is displayed in red and it includes contributions from the SIDM halo (blue solid), stars (magenta dashed), and gas (magenta dot-dashed). The predictions of the corresponding CDM halo (dotted) and the SIDM halo neglecting the influence of the baryons (asterisk) are also shown. The UGC 5721 and UGC 5750 fits are different from those in the main text: they are less extreme by $0.5\sigma$ and demonstrate the mild degeneracy between $c_{200}$ and $M_{200}$. We also provide two fits each for UGC 3371 and IC 2574 to demonstrate the effects of changing $c_{200}$. In our analysis, we do not take into account the influence of the gas component on the isothermal dark matter profile, which could be moderately important for the fit to IC 2574. }
\end{figure*}

\begin{figure*}[htb]
\centering
\begin{tabular}{@{}cc@{}}
\includegraphics[scale=0.7]{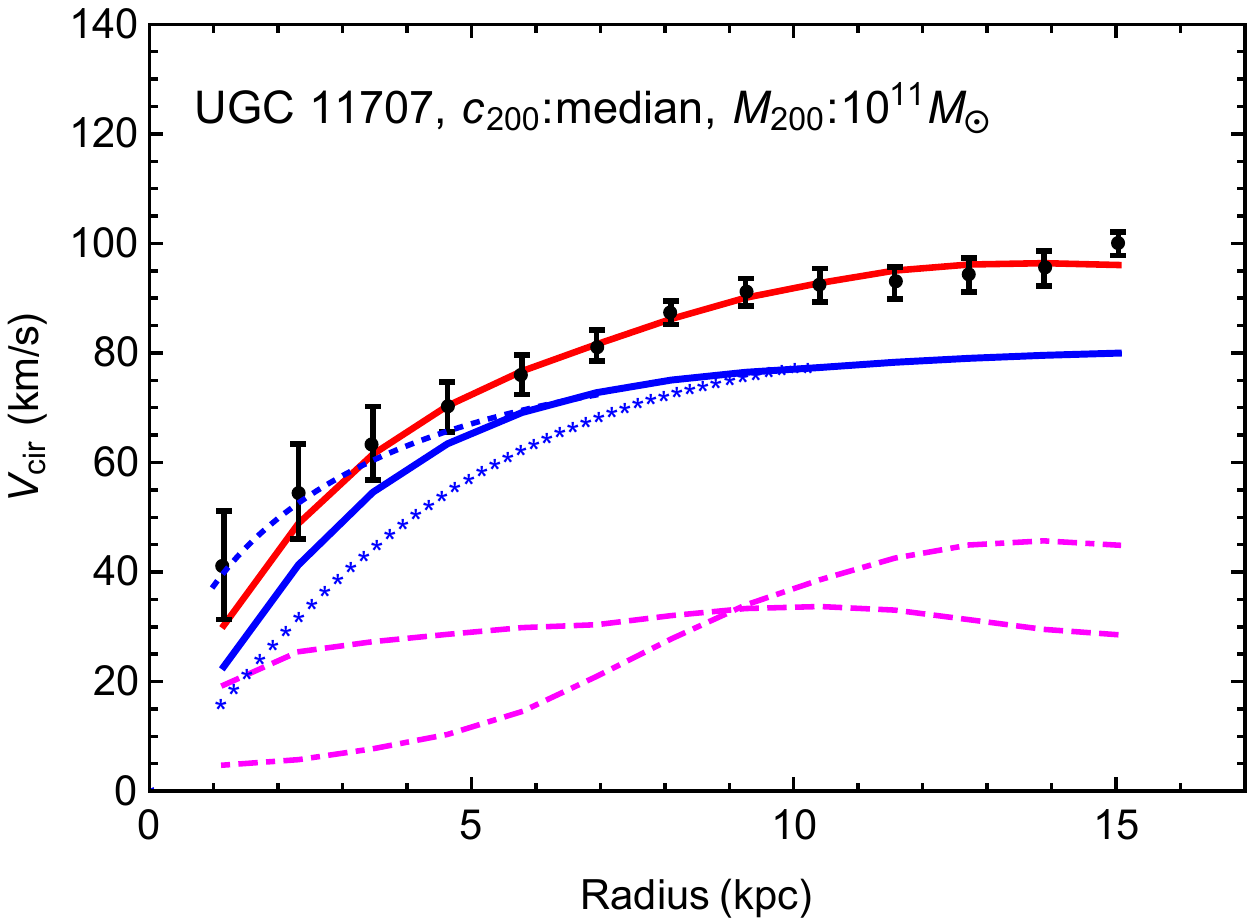}&
\includegraphics[scale=0.7]{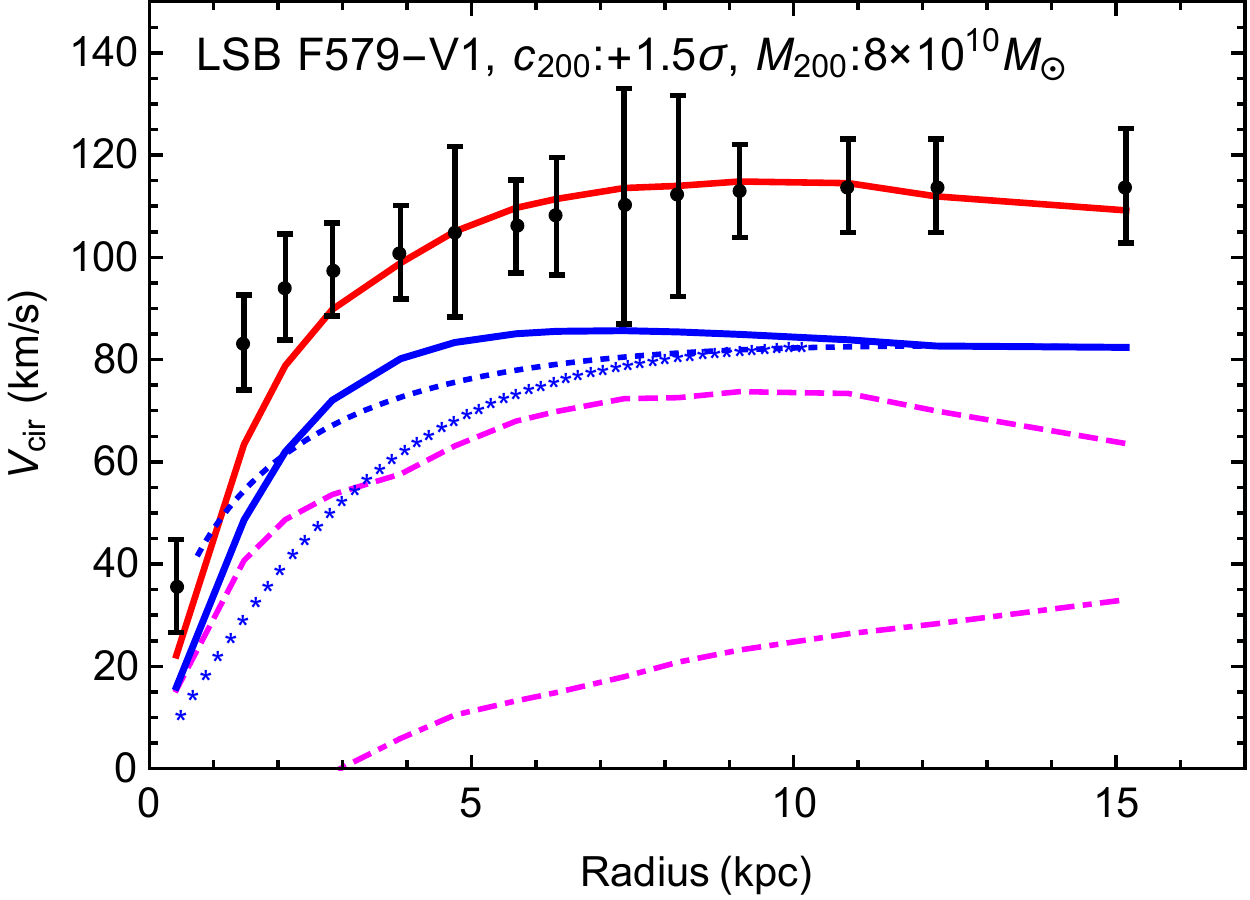} \\
\includegraphics[scale=0.7]{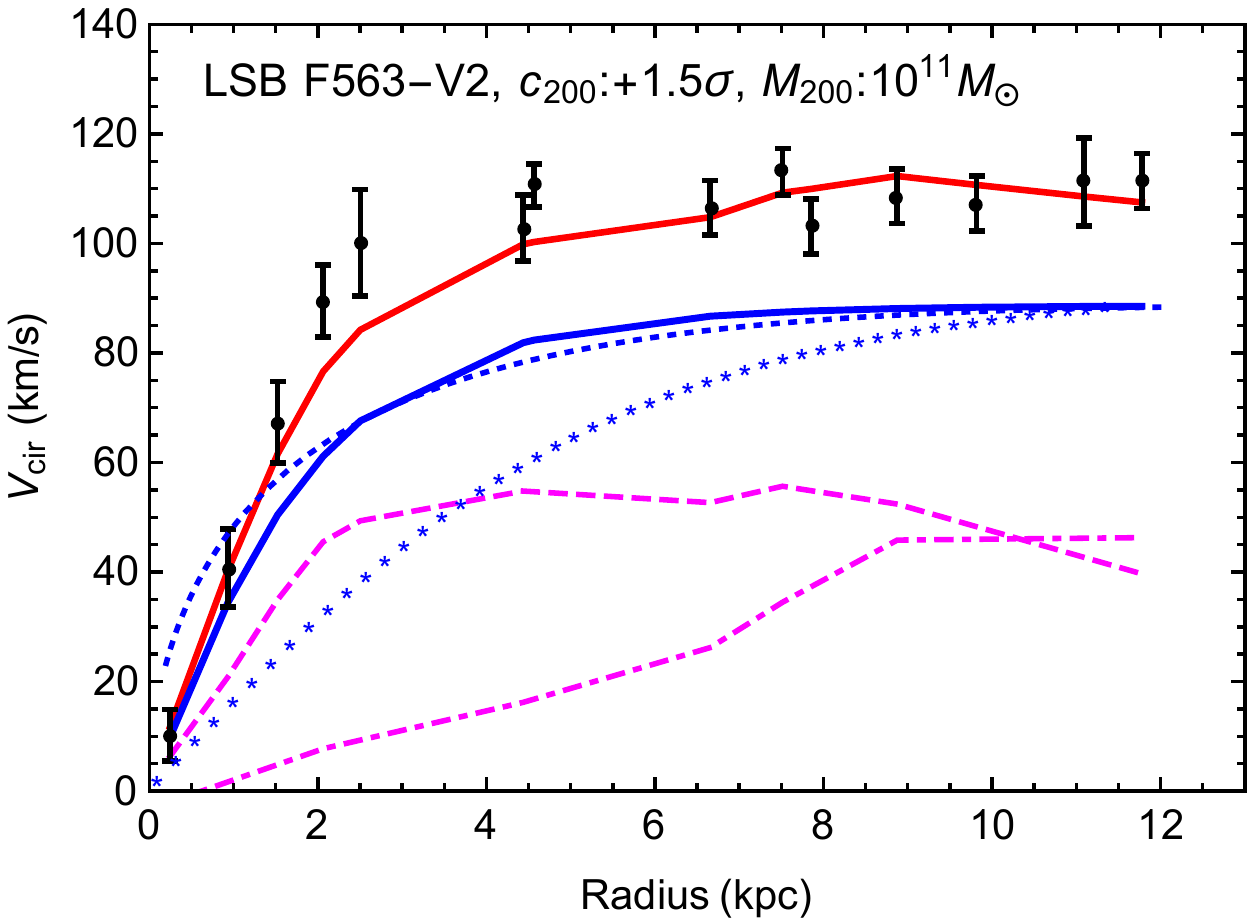}&
\includegraphics[scale=0.7]{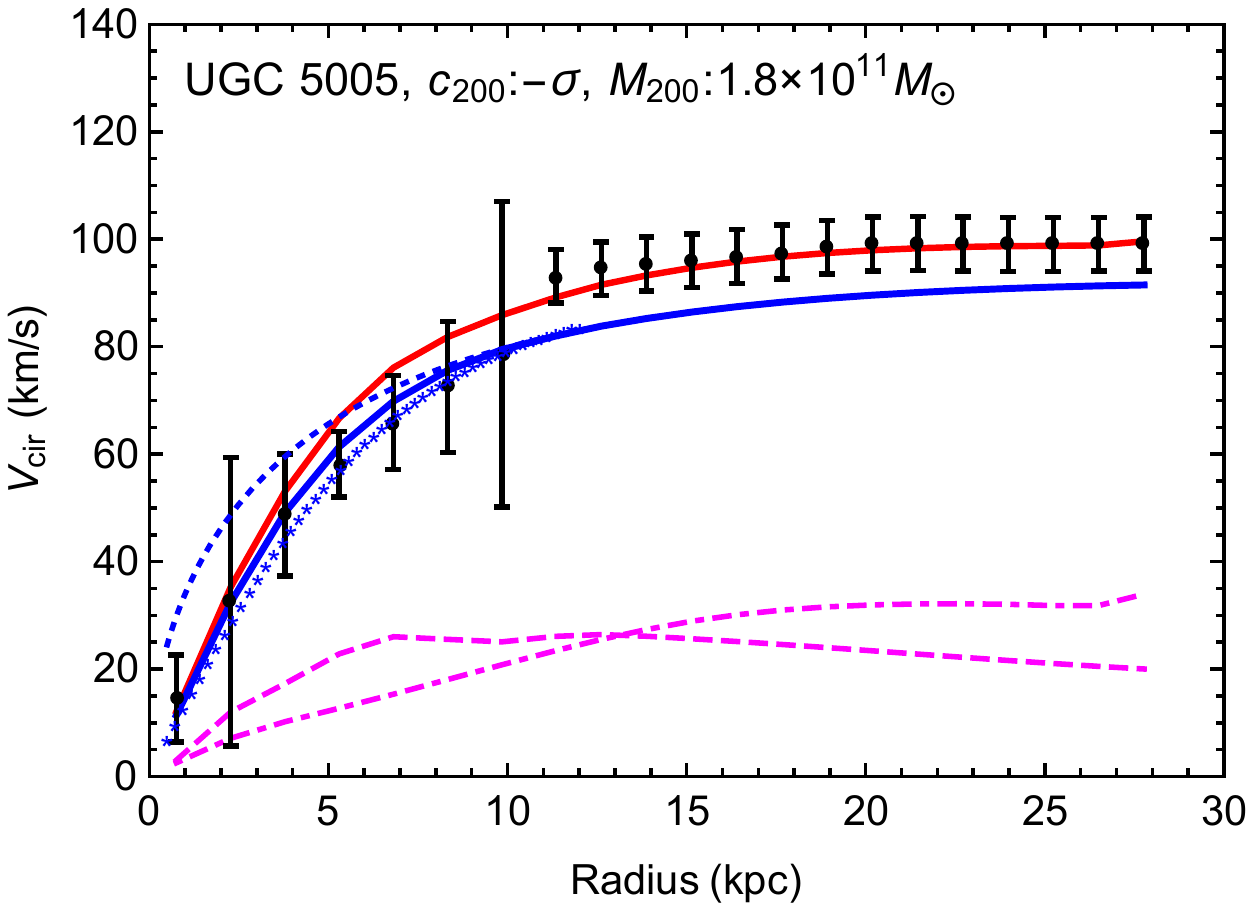}
\end{tabular}
\caption{\label{fig:v100} Galaxies with $V_f\approx100~{\rm km/s}$.  Data for UGC 11707, F579-V1, F563-V2, and UGC 5005 from \cite{swaters2009_whisp, swaters2003},  \cite{deblok1996, mcgaugh2001}, \cite{deblok1996, swaters2003, kuziodenaray2006}, and \cite{vanderhulst1993, deBlok:2002tg} respectively. The total fit is displayed in red and it includes contributions from the SIDM halo (blue solid), stars (magenta dashed), and gas (magenta dot-dashed). The predictions of the corresponding CDM halo (dotted) and the SIDM halo neglecting the influence of the baryons (asterisk) are also shown.}
\end{figure*}

\begin{figure*}[htb]
\centering
\begin{tabular}{@{}cc@{}}
\includegraphics[scale=0.7]{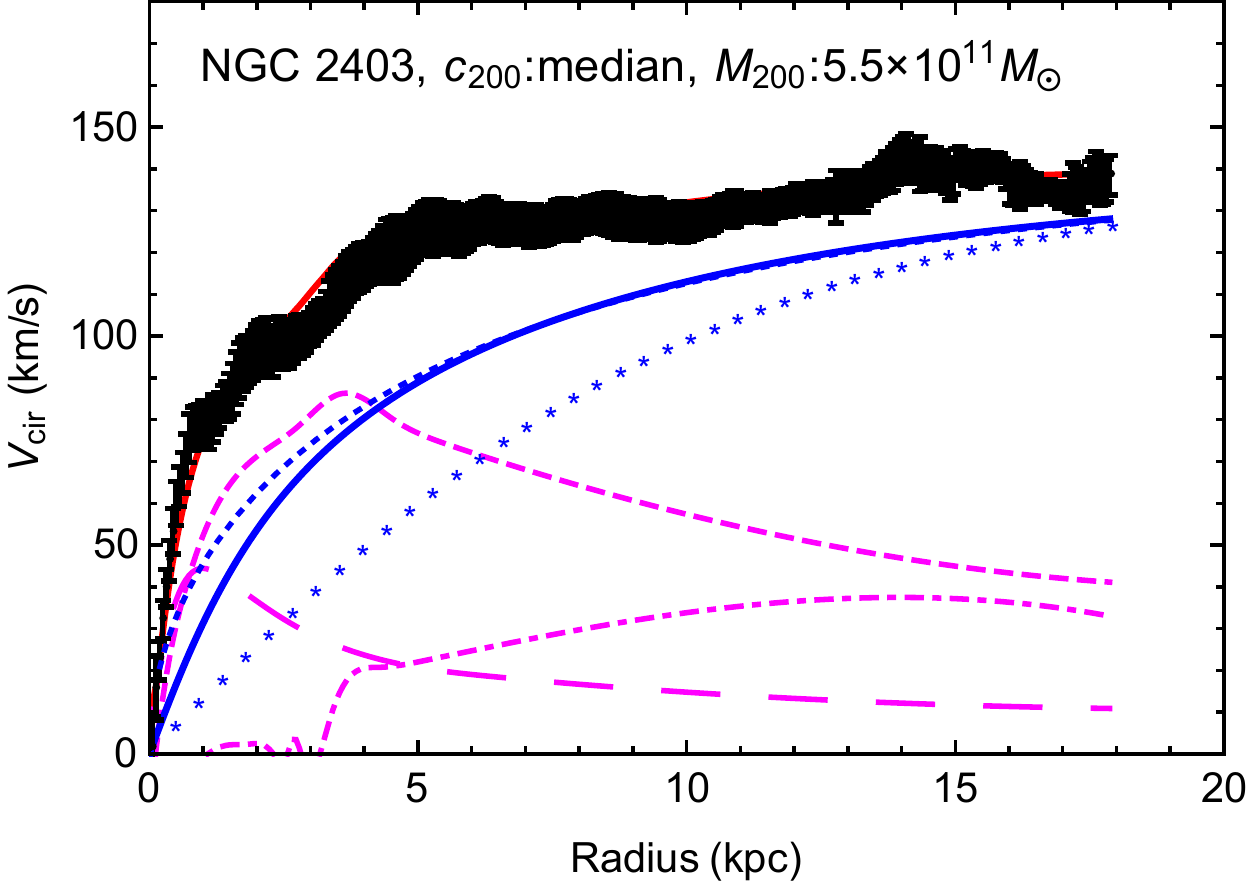}&
\includegraphics[scale=0.7]{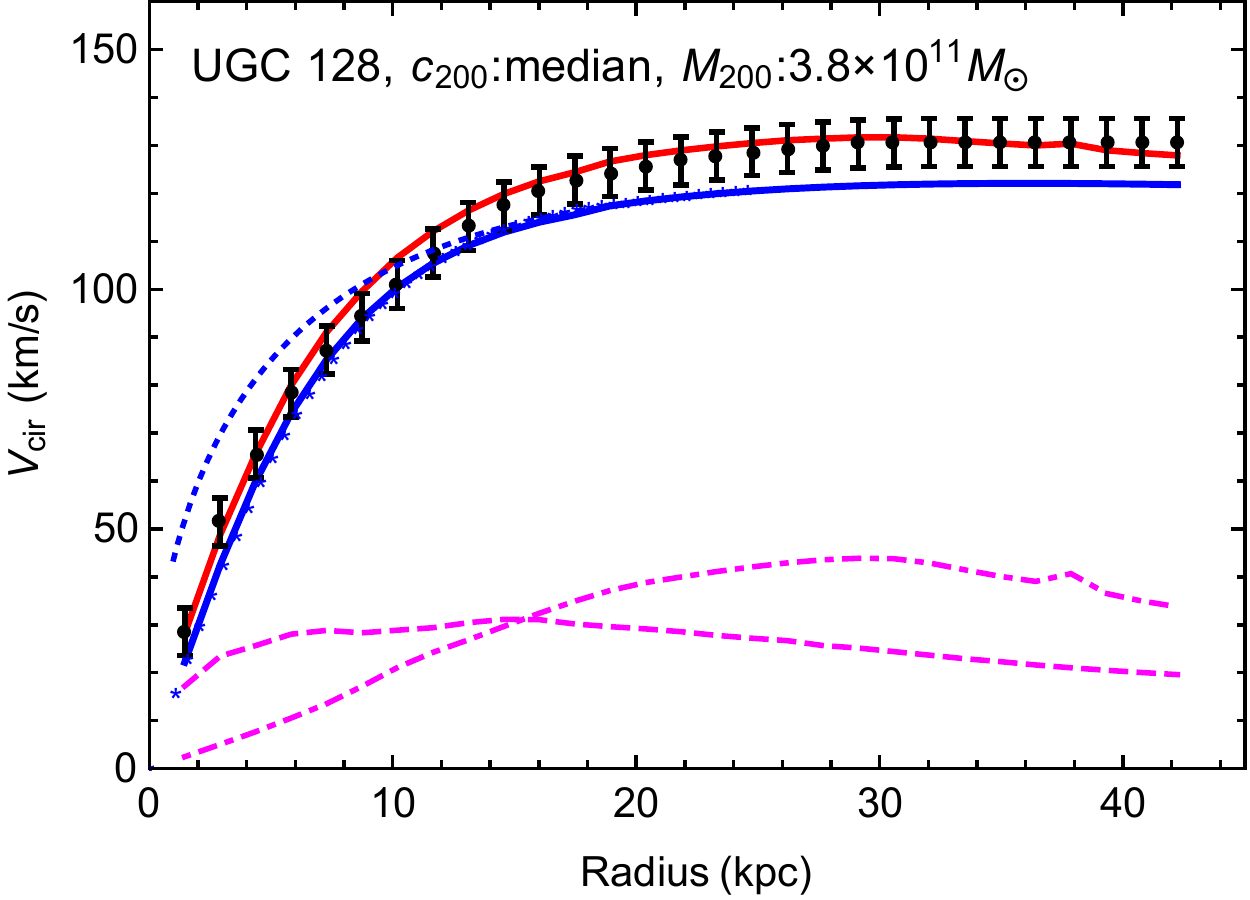} \\
\includegraphics[scale=0.7]{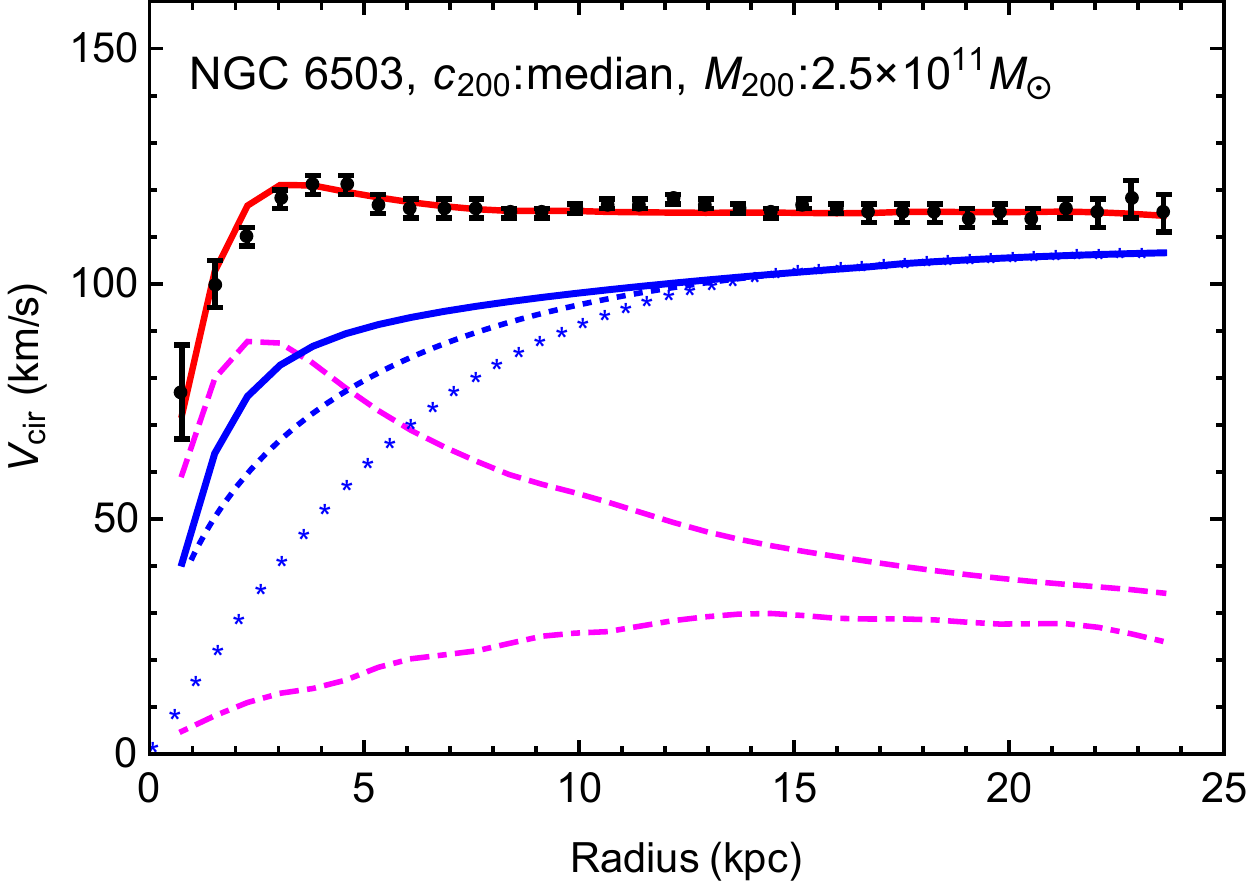}&
\includegraphics[scale=0.7]{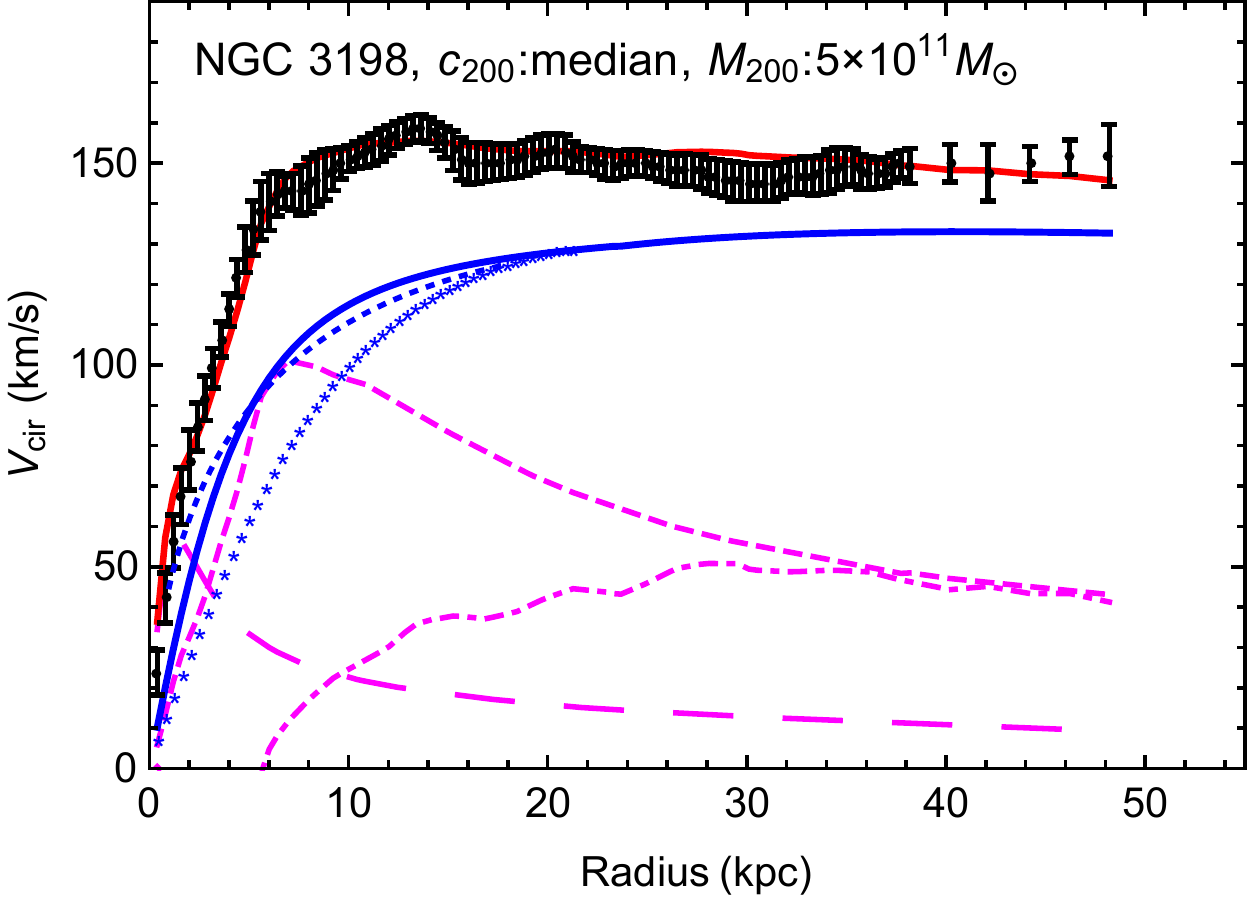}\\
\includegraphics[scale=0.7]{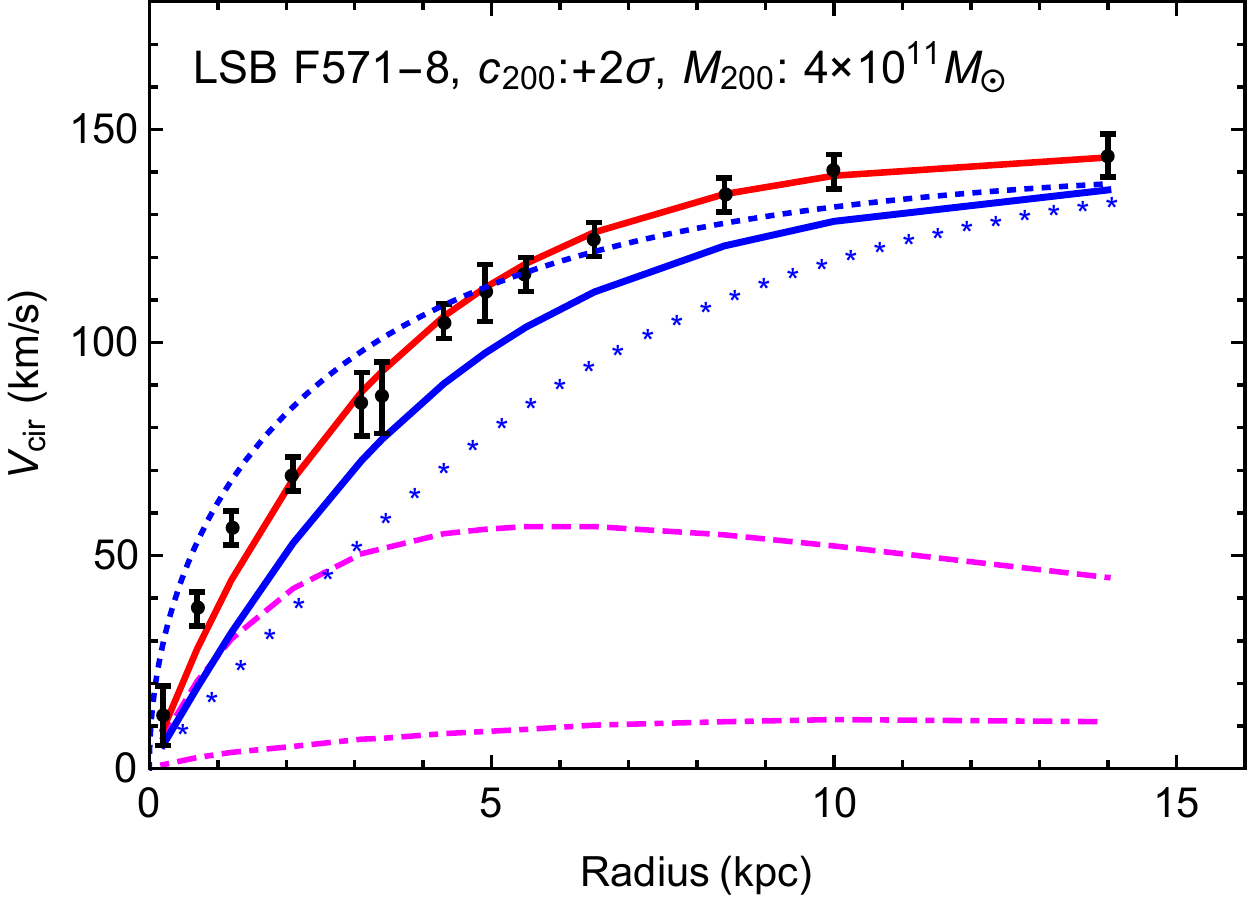}&
\includegraphics[scale=0.7]{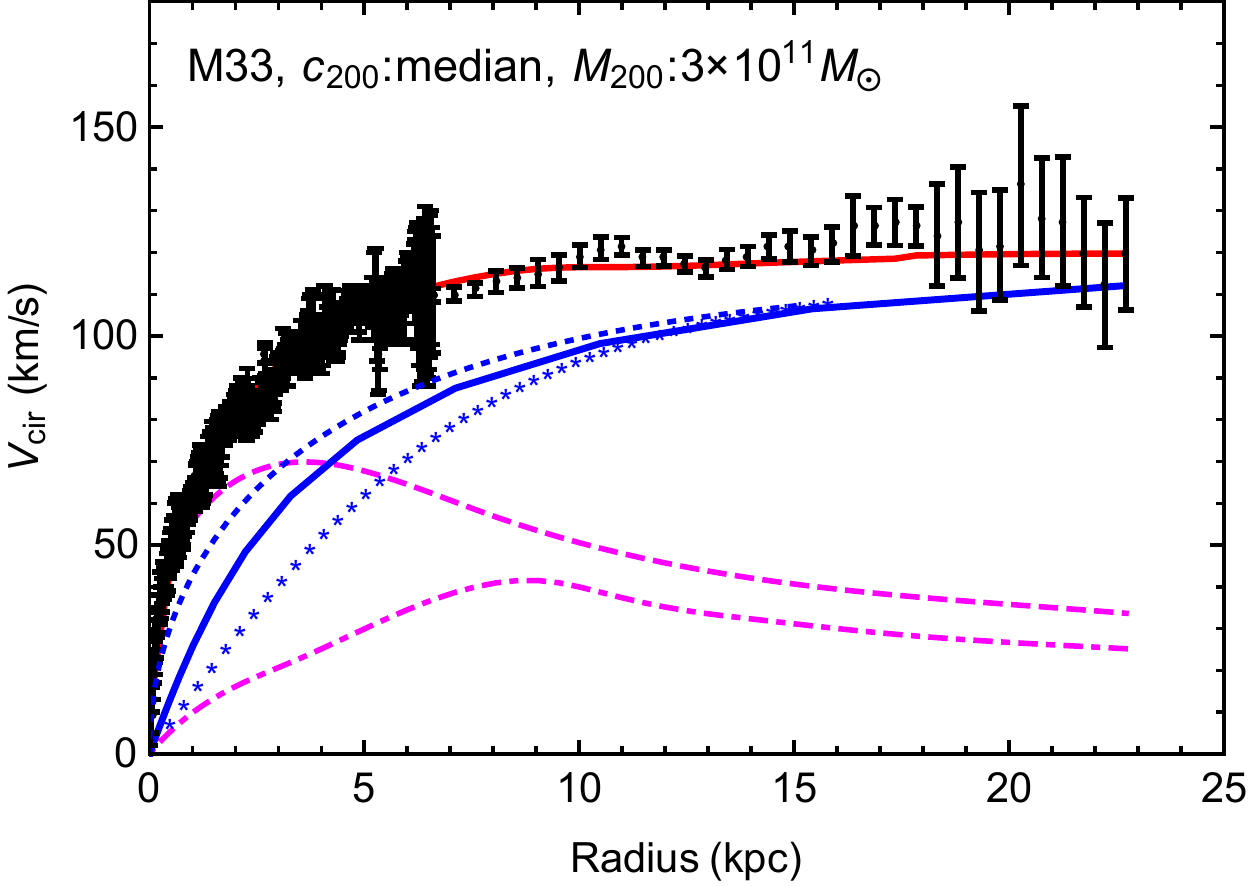}\\
\end{tabular}
\caption{\label{fig:v120} Galaxies with $V_f\approx120\textup{-}150~{\rm km/s}$. Data for NGC 2403, UGC 128, NGC 6503, NGC 3198, and LSB F571-8 from \cite{deblok2008}, \cite{vanderhulst1993}, \cite{begeman1987},  \cite{deblok2008, gentile2013}, \cite{mcgaugh2001}, and \cite{Corbelli2014, Kam2015} respectively. The total fit is displayed in red and it includes contributions from the SIDM halo (blue solid), stars (magenta dashed), and gas (magenta dot-dashed). The predictions of the corresponding CDM halo (dotted) and the SIDM halo neglecting the influence of the baryons (asterisk) are also shown.}
\end{figure*}

\begin{figure*}[htb]
\centering
\begin{tabular}{@{}cc@{}}
\includegraphics[scale=0.7]{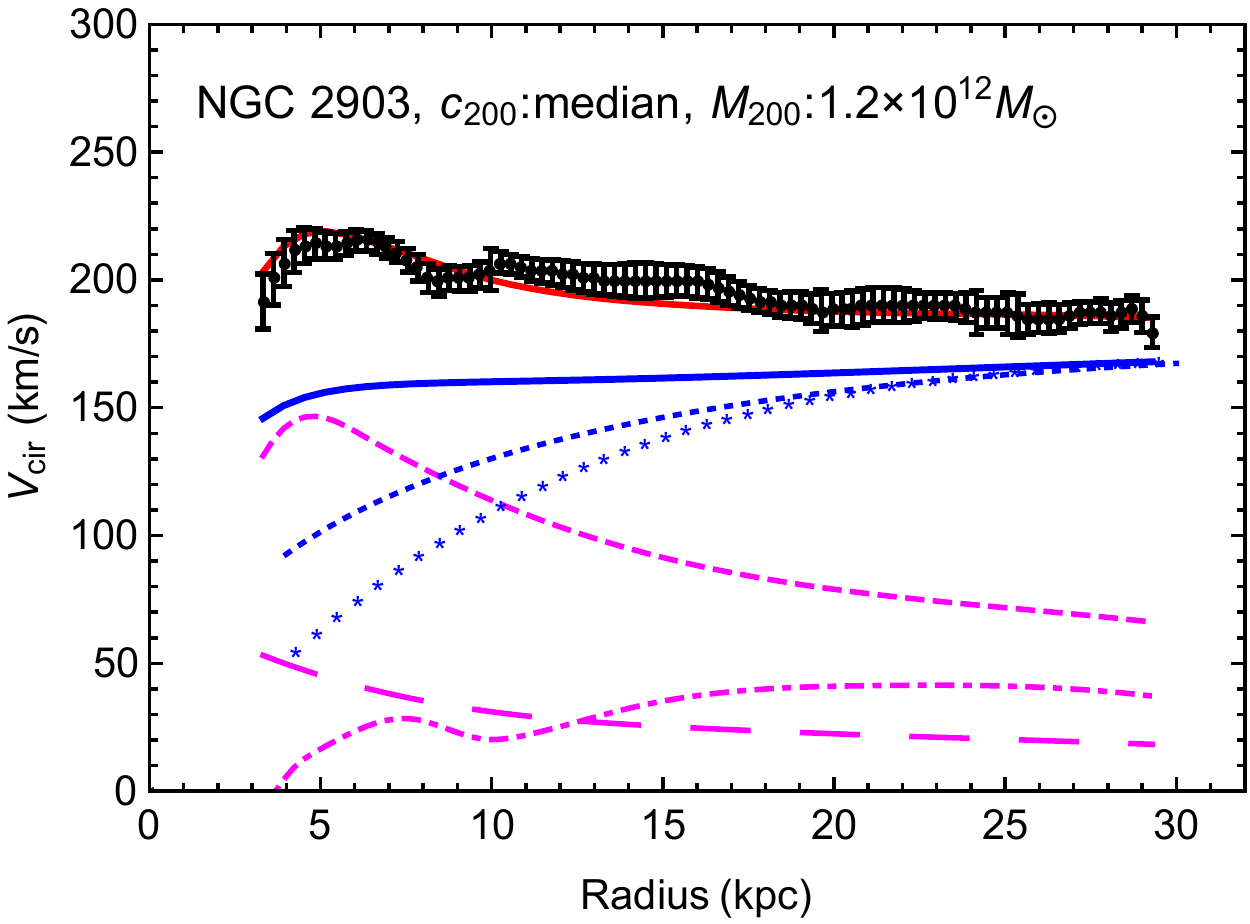}&
\includegraphics[scale=0.7]{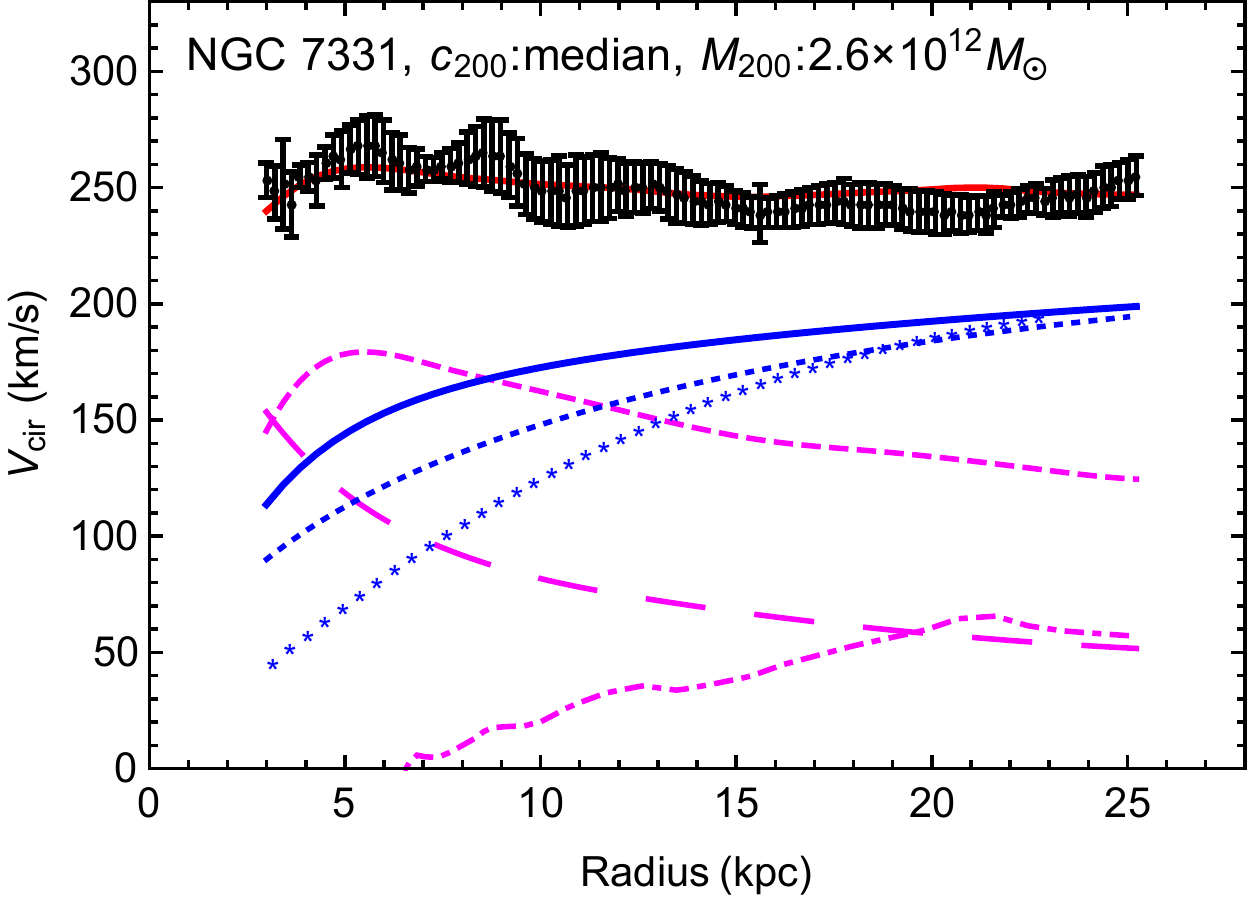} \\
\includegraphics[scale=0.7]{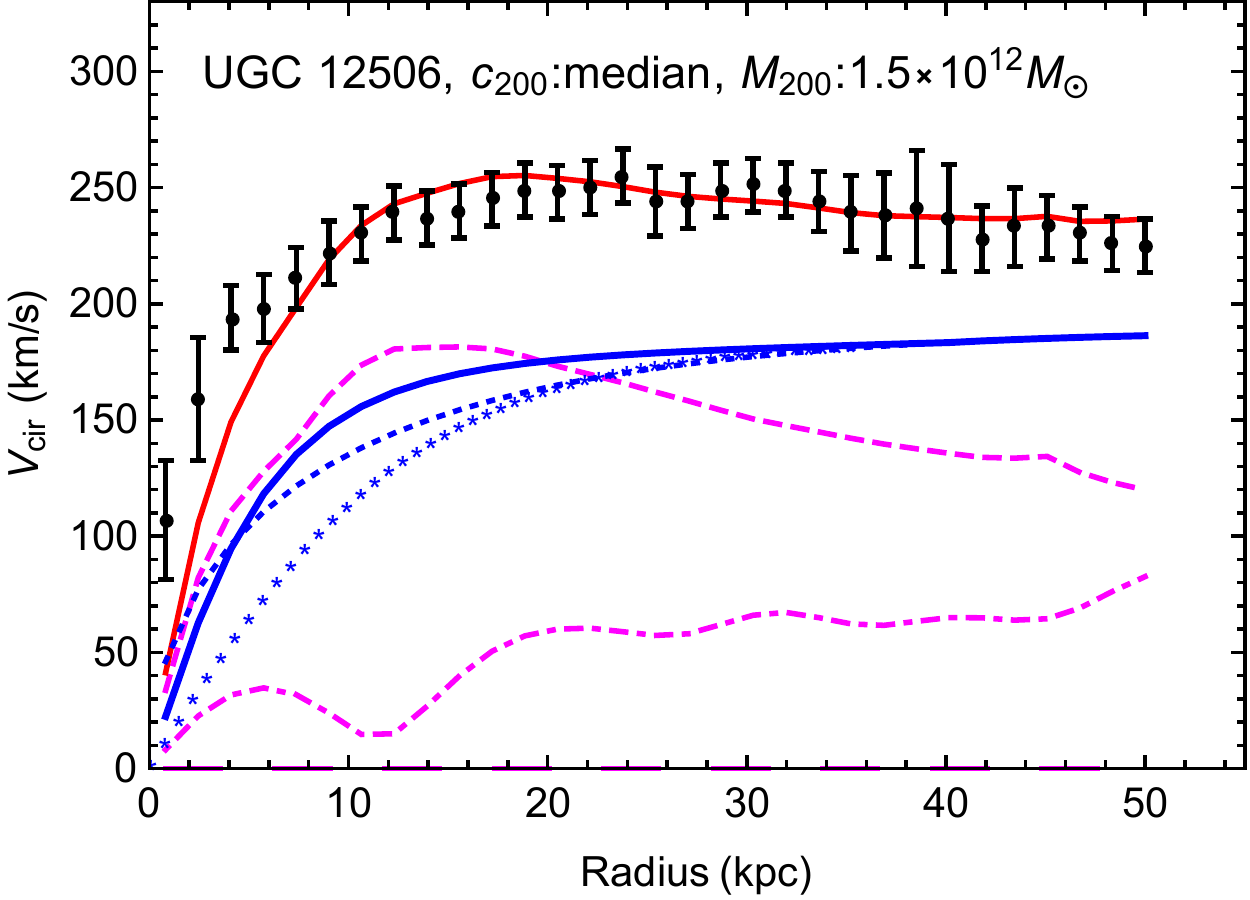}&
\includegraphics[scale=0.7]{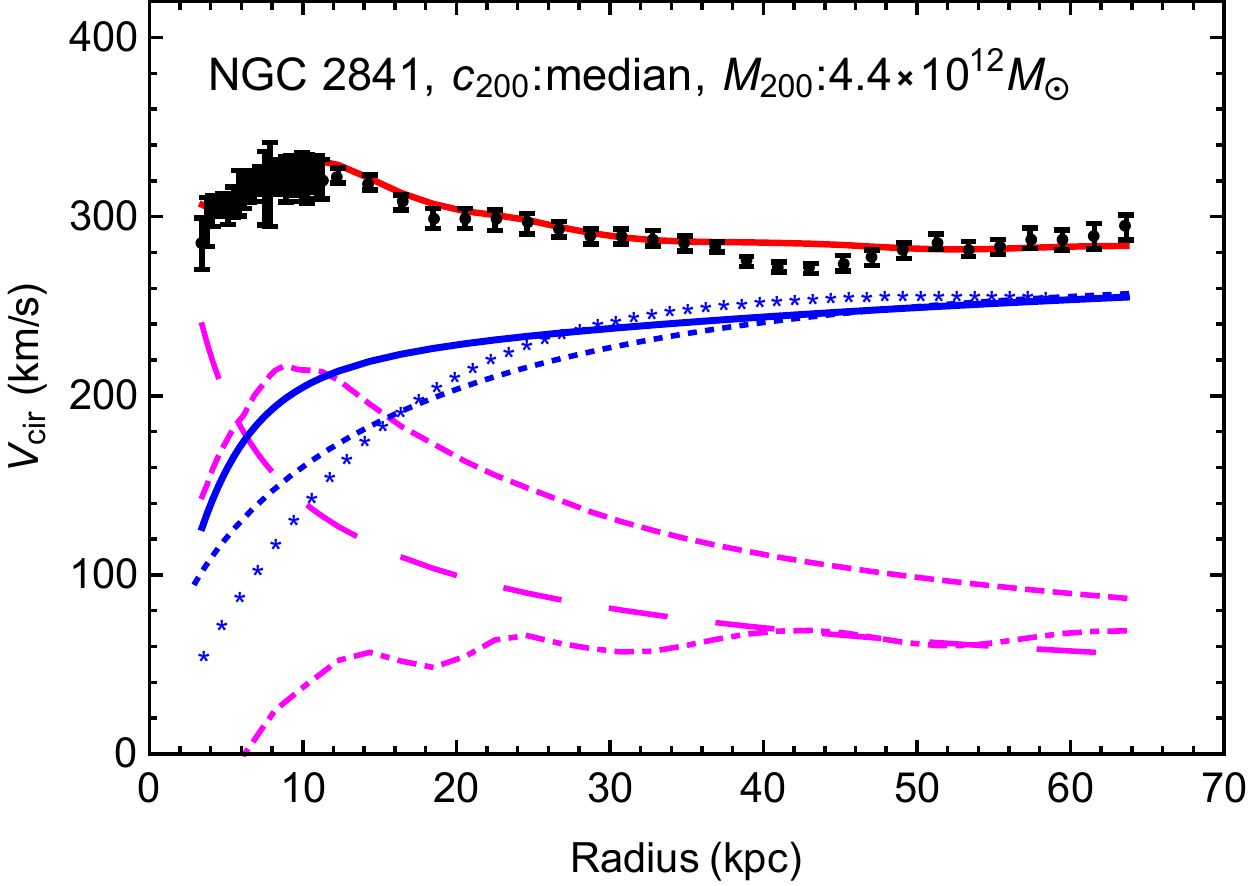} \\
\end{tabular}
\caption{\label{fig:v250} Galaxies with $V_f\approx200\textup{-}300~{\rm km/s}$. Data for NGC 2903 and NGC 7331 from \cite{deblok2008}, UGC 12506 and NGC 2841 from \cite{Lelli2016_sparc}. The total fit is displayed in red and it includes contributions from the SIDM halo (blue solid), stars (magenta dashed), gas (magenta dot-dashed), and bulge (magenta long-dashed). The predictions of the corresponding CDM halo (dotted) and the SIDM halo neglecting the influence of the baryons (asterisk) are also shown.}
\end{figure*}

\end{document}